\documentclass[aps,reprint,longbibliography,superscriptaddress]{revtex4-1}
\usepackage{amsmath}
\usepackage{amssymb}
\usepackage{bm}
\usepackage{epsfig}
\usepackage{graphicx}
\usepackage[dvipsnames]{xcolor}
\usepackage{color}
\usepackage[unicode=true,colorlinks=true,citecolor=blue,urlcolor=blue]{hyperref}
\usepackage[T2A]{fontenc}
\usepackage[cp1251]{inputenc}
\usepackage[english]{babel}
\usepackage{braket}

\newcommand{\sign}{\mathop{\rm sign}\nolimits}

\newcommand{\Tr}{\mathop{\rm Tr}\nolimits}

\newcommand{\erfc}{\mathop{\rm erfc}\nolimits}

\renewcommand{\sinh}{\mathop{\rm sh}}
\renewcommand{\cosh}{\mathop{\rm ch}}

\newcommand{\e}{\mathrm{e}}

\renewcommand{\i}{{\rm i}}
\renewcommand{\d}{\mathrm d}
\renewcommand{\emph}{\textit}

\renewcommand{\braket}[1]{\left\langle #1 \right\rangle}
\newcommand{\Mn}{Mn$^{2+}$}

\newcommand{\addDima}[1]{#1}
\newcommand{\commentDima}[1]{}%\textcolor{red}{\textbf{Dima: }\textit{#1}}}
\begin{document}

\title{Optical Resonance Shift Spin Noise Spectroscopy}
% \title{Indirect optical spectroscopy of quantum spin noise}

\author{D.~S.~Smirnov}
\email{smirnov@mail.ioffe.ru}
\affiliation{Ioffe Institute, 194021 St. Petersburg, Russia}
\author{K.~V.~Kavokin}
\affiliation{Spin Optics Laboratory, Saint Petersburg State University, 198504 St. Peterbsurg, Russia}

%\date{\today}

\begin{abstract}
Quantum spin fluctuations provide a unique way to study spin dynamics without system perturbation. Here we put forward an optical resonance shift spin noise spectroscopy as a powerful tool to measure the spin noise of various systems from magnetic impurities in solids to free atoms and molecules.
The quantum spin fluctuations in these systems can shift the optical resonances by more than the homogeneous linewidth and produce huge Faraday rotation noise. We demonstrate, that the resonance shift spin noise spectroscopy \addDima{gives access} to the high order spin correlators, which contain complete information about the spin dynamics in contrast with the second order correlator measured by conventional Pauli-blocking spin noise spectroscopy. The high order quantum spin correlators manifest themselves as a comb of peaks in the Faraday rotation noise spectra in transverse magnetic field.
This effect is closely related with the multispin flip Raman scattering observed in the Mn-doped nanostructures.
% Quantum spin fluctuations provide a unique way to study spin dynamics in solids without \addDima{its} perturbation by means of the spin noise spectroscopy. Experimental measurement of the correlation functions of electron and hole spins  beyond the second order is essentially impossible because of the weak spin signals. Here we put forward \addDima{an optical resonance shift spin} noise spectroscopy as a tool to measure the high order spin correlators of magnetic impurities. Their quantum spin fluctuations shift the optical resonance energies by more than the homogeneous line width, which produces the huge noise signals. We predict the comb of the peaks in the Faraday rotation noise spectra, which are produced by the spin correlations of high orders. We demonstrate the relation of this effect with the multispin flip Raman scattering observed in the Mn-doped nanostructures and demonstrate the possibility to \addDima{measure} quantum spin noise of various systems including spins of the host lattice and impurities as well as the spins of the free atoms and molecules.
\end{abstract}

\maketitle

%====================================================

\section{Introduction}

The quantum spin fluctuations were first predicted by Felix Bloch back in 1946~\cite{Bloch46}. With the development of the experimental techniques, the optical spin noise spectroscopy appeared and eventually became a powerful tool for the spin dynamics investigation in a broad class of paramagnetic media, from atomic gases to semiconductors~\cite{Zapasskii:13,Oestreich-review}. In typical experiments, the spin fluctuations within the small volume of a paramagnetic material produce a stochastic Faraday rotation of the linearly polarized light, which probes the system, and the spin noise spectra are obtained by the Fourier transformation of the time-dependent Faraday rotation.

In a magnetic field perpendicular to the probe beam, the spin noise spectrum shows peaks at the Larmor frequencies of the studied spins, similar to the optically detected magnetic resonance.
%In the semi-classical consideration, these resonances result from precession of spin fluctuations in the applied magnetic field, which induces modulation of refraction indices. %It has been shown theoretically~\cite{Gorbovitskii} (see also Ref.~\cite{NonlinearSNS}) that the
Quantum-mechanically the spin noise signal can be considered as a result of the interference of the probe beam with the light emission caused by spin-flip forward scattering of the probe light~\cite{Gorbovitskii,NonlinearSNS}.

In atomic gases, the dependence of the light scattering amplitude on the spin state of an atom is provided by the Pauli blocking of the optical transitions in certain polarizations, defined by the probed spins orientation. This scenario is also realized for electrons and holes in semiconductors with the pronounced spin-orbit interaction, such as GaAs or \addDima{CdTe~\cite{ivchenko05a}.} In the charged quantum dots (QDs), for example, for the electron spin-up or spin-down state the optical transitions to the singlet heavy hole trion state are possible for $\sigma^+$ or $\sigma^-$-polarized light \addDima{only}, respectively. In thermal equilibrium, the number of spin-up and spin-down electrons is the same on average, but stochastic spin fluctuations produce a weak Gaussian Faraday rotation noise, which is measured. This type of experiments can be called ``Pauli-blocking \addDima{spin noise} spectroscopy''.

The small Faraday rotation angles in Pauli-blocking spin noise spectroscopy make it difficult to detect the spin correlation functions of the orders higher than two~\cite{PhysRevA.93.033814,4order_exp}. In the same time, the complete information about the spin dynamics including its intrinsic quantum properties can be obtained from the complete set of the spin correlators of all orders \addDima{only}~\cite{Liu2010,Bednorz,PhysRevB.98.205143}. As a minimal extension of the standard theories the weak measurements of the third and fourth order spin correlators were described~\cite{PhysRevB.90.205419,SNS_Universality}.% Despite the natural assumption, that for large systems the spin noise is Gaussian, this was never checked experimentally, and the measurement of high order spin correlators remains highly desired.}

An alternative connection between the spin system and the light polarization is realized in diluted magnetic semiconductors~\cite{Kossut_book}. In this case the probed spins belong to the $d$-shell electrons of Mn$^{2+}$ ions, embedded into the crystal lattice. Mn atoms do not create localized charge carrier states, and their spins do not affect interband optical transitions directly. However, their spins are coupled to the spins of the conduction electrons and holes by the $sp-d$ exchange interaction. The corresponding coupling constant is very large, of the order of 1 eV. Due to this interaction, fluctuations of the magnetic-ion spins modulate the energies of the interband optical transitions (most often involving localized excitons), and this creates a polarization noise of the probe light, so we call this type of experiments ``resonance shift spin noise spectroscopy''. Atomic-like hyperfine structure of {\Mn} spin levels was resolved in such experiments with the very diluted $\mbox{CdMnTe}$ quantum wells in weak magnetic fields~\cite{Scalbert2015}. The same mechanism is responsible for the observation of the nuclear spin noise in GaAs~\cite{NuclearNoise,book_Glazov}. Moreover the resonance shift spin noise spectroscopy can be applied to any impurities in the semiconductors or to the spins of nuclei of free atoms and molecules.

The general arguments, which followed the first Pauli-blocking spin noise measurement~\cite{Gorbovitskii} establish the relation between the spin noise spectrum and the spin flip Raman spectrum. In the case of the resonance shift spectroscopy, this relation apparently breaks down. Indeed, there \addDima{exists} a phenomenon of the multispin flip scattering, when the resonant Raman spectrum of Mn doped nanostructures shows a comb of up to 15 equally spaced peaks~\cite{Stuhler1995,Kozyrev2019}.  These spectra are explained by the scattering via the virtual magnetic-polaron states, and the observed phenomenon is therefore essentially \addDima{quantum~\cite{Stuhler1996,ClassicallyLarge}}. On the other hand, the Larmor precession of the spin fluctuation of one, several or many {\Mn} ions induces the peak in the spin noise spectrum at the single Larmor frequency only, \addDima{and} not at its multiplies.

% the multispin flip Raman spectrum is completely different from the spin noise spectrum. The resonant Raman spectra of nanostructures moderately doped with Mn in the transverse magnetic field (Voigt geometry) consists of many (up to 15) equally spaced peaks~\cite{Stuhler1995,Kozyrev2019}. These peaks result from the simultaneous flips of the corresponding number of Mn$^{2+}$ spins.  These spectra are explained by scattering via the virtual magnetic-polaron states, and the observed phenomenon is therefore essentially quantum~\cite{Stuhler1995,Stuhler1996,ClassicallyLarge}. In the same time, the spin noise spectrum consists of only one peak at the Larmor spin precession frequency, which apparently violates the relation between the Raman the spin spin noise spectra.

In this work, we show that the multispin flip Raman scattering is a counterpart of the high order quantum spin noise spectra, which can be observed by means of resonance shift spin spectroscopy in semi-magnetic structures. We develop a general theory of the resonance shift quantum spin noise spectroscopy and describe in a unified way the spin noise and Raman spectra. We demonstrate, that the shape of the spectra is different for the thermal and quantum spin noise. Detection of the high order spin correlators allows one to completely describe the spin dynamics, and to distinguish between Gaussian and non-normal spin fluctuations. In particular, we show that for deep impurities in semiconductors and in atomic systems the spin noise spectra strongly differ from the Gaussian noise spectra.

The paper is organized as follows: In Sec.~\ref{sec:model} we present a model and derive the general expression for the Faraday rotation noise spectrum in the framework of the resonance shift spin noise spectroscopy. In what \addDima{follows,} we focus on the semimagnetic quantum wells and QDs, where we anticipate the fastest experimental measurement of the higher order spin correlators. In Sec.~\ref{sec:general} we establish the relation between the Faraday rotation noise and Raman spin-flip spectra in different polarizations. In Sec.~\ref{sec:indirect_spectra} we calculate and describe the spectra for the Gaussian spin noise and in Sec.~\ref{sec:abnormal} we describe the non-normal spin noise. Finally, we discuss the applications of our theory to the different spin systems from the solid state to free atoms and summarize our findings in Sec.~\ref{sec:conclusion}.

%====================================================
\section{Model}
\label{sec:model}
%\section{Spin-dependent exciton dipole polarization}

As a model system for the resonance shift spin noise spectroscopy we consider a II-VI semiconductor doped with manganese. The spins of Mn$^{2+}$ atoms can be optically monitored via a localized exciton resonance. We assume, that the excitons are localized \addDima{at defects, in} QDs, or at imperfections of the interfaces of a quantum well. The general form of the Hamiltonian is
\begin{equation}
  \label{eq:H}
  \mathcal H(t)=\mathcal H_0+\mathcal H_{exc}+\mathcal H_{int}+\mathcal V(t).
\end{equation}
Here $\mathcal H_0$ is the Hamiltonian of {\Mn} spin system, $\mathcal H_{exc}$ is the exciton Hamiltonian, $\mathcal H_{int}$ describes the interaction between exciton and {\Mn} spins, and $\mathcal V(t)$ stands for the coherent optical excitation of excitons.
%Generally, $\mathcal H_0$ can also include the interaction  interaction with external magnetic fields and with the environment.
%The Hamiltonian $\mathcal H_{exc}$ describes the whole fine structure of the excitonic levels, and includes the electron-hole exchange interaction. 

The spin dependent part of the interaction of magnetic atoms with excitons stems from the exchange interaction with electron and hole in \addDima{the} exciton. The general form of this interaction is~\cite{Merkulov_book}
\begin{equation}
  \label{eq:H_int}
  \mathcal H_{int}=\hbar\sum_i\left[\omega_{ex,i}^e \bm S^e\bm I_{i}+\sum_\alpha\omega_{ex,i}^{h,\alpha} S_\alpha^hI_{i,\alpha}\right],
\end{equation}
where $i$ enumerates {\Mn} spins $\bm I_i$, $\bm S^e$ and $\bm S^h$ are the electron and hole spins in the given exciton, respectively, $\omega_{ex,i}^e$ and $\omega_{ex,i}^{h,\alpha}$ are the corresponding exchange interaction constants with \addDima{the Cartesian index $\alpha=x,y,z$}. Due to the different symmetry of the electron and hole Bloch wave functions in the $\Gamma$ valley, the electron exchange interaction is isotropic, while for the heavy hole it is not. This anisotropy plays an important role for this system, and has the same origin as the anisotropy of the effective $g$-factor.
% Similarly to the hyperfine interaction with the host lattice nuclei, this anisotropy is defined by the heavy hole Bloch wave function.

We consider the optical detection of {\Mn} spins by resonant laser probe light, which is described by the term
\begin{equation}
  \label{eq:V}
  \addDima{\mathcal V}=-\bm P^\dag\bm E\e^{-\i\omega_p t}  + \rm{H.c.},
\end{equation}
%\commentDima{The sign here was wrong. I have to check all the signs below!}
in the Hamiltonian~\eqref{eq:H}. Here $\bm P$ is the dipole moment operator (in Schr\"odinger representation), $\omega_p$ is the probe frequency and $\bm E$ is the amplitude of the incident electric field. The probe light induces an exciton dipole polarization, which is described by the Heisenberg time dependent operator $\bm P(t)$. In Appendix~\ref{app:P} we demonstrate, that it has the form $\bm P(t)=\bm P \psi(t)$, where $\psi(t)$ is a dimensionless operator, which satisfies the \addDima{transparent} equation
\begin{multline}
  \label{eq:psi}
  \frac{\d\psi(t)}{\d t}=\frac{\i}{\hbar}\bm P^\dag \bm E\e^{-\i\omega_p t}-\frac{\i}{\hbar}\left[\mathcal H_{exc}+\tilde{\mathcal H}_{int}(t)\right]\psi(t)\\-\gamma\psi(t).
\end{multline}
Here we introduced the interaction Hamiltonian in the interaction representation
\begin{equation}
  \label{eq:Hint_int}
  \tilde{\mathcal H}_{int}(t)=\e^{\i\mathcal H_0 t/\hbar}\mathcal H_{int}\e^{-\i\mathcal H_0 t/\hbar}
\end{equation}
and an optical transition dephasing rate $\gamma$. The equation~\eqref{eq:psi} can be formally integrated, and the result for the time dependent exciton polarization reads
\begin{multline}
  \label{eq:P_gen}
  \bm P(t)=\frac{\i}{\hbar}\bm P\int\limits_0^\infty\e^{-\i\omega_p(t-\tau)-\gamma\tau}\\
  \times\mathcal T\exp\left[-\frac{\i}{\hbar}\int\limits_0^\tau \mathcal H_{exc}'(t-\tau')\d\tau'\right]\d\tau(\bm P^\dag\bm E).
\end{multline}
Here $\mathcal T\exp$ denotes the normal time ordered exponential  \addDima{(later times on the left)} and $\mathcal H_{exc}'(t)=\mathcal H_{exc}+\tilde{\mathcal H}_{int}(t)$ is an effective time dependent exciton Hamiltonian. It is this part of the expression that contains information about parameters of {\Mn} spin dynamics ($\mathcal H_0$). The obtained general expression allows one to describe various physical systems and experimental conditions. Below we consider a specific case, when this expression is greatly simplified.

\begin{figure}
  \centering
  \includegraphics[width=0.9\linewidth]{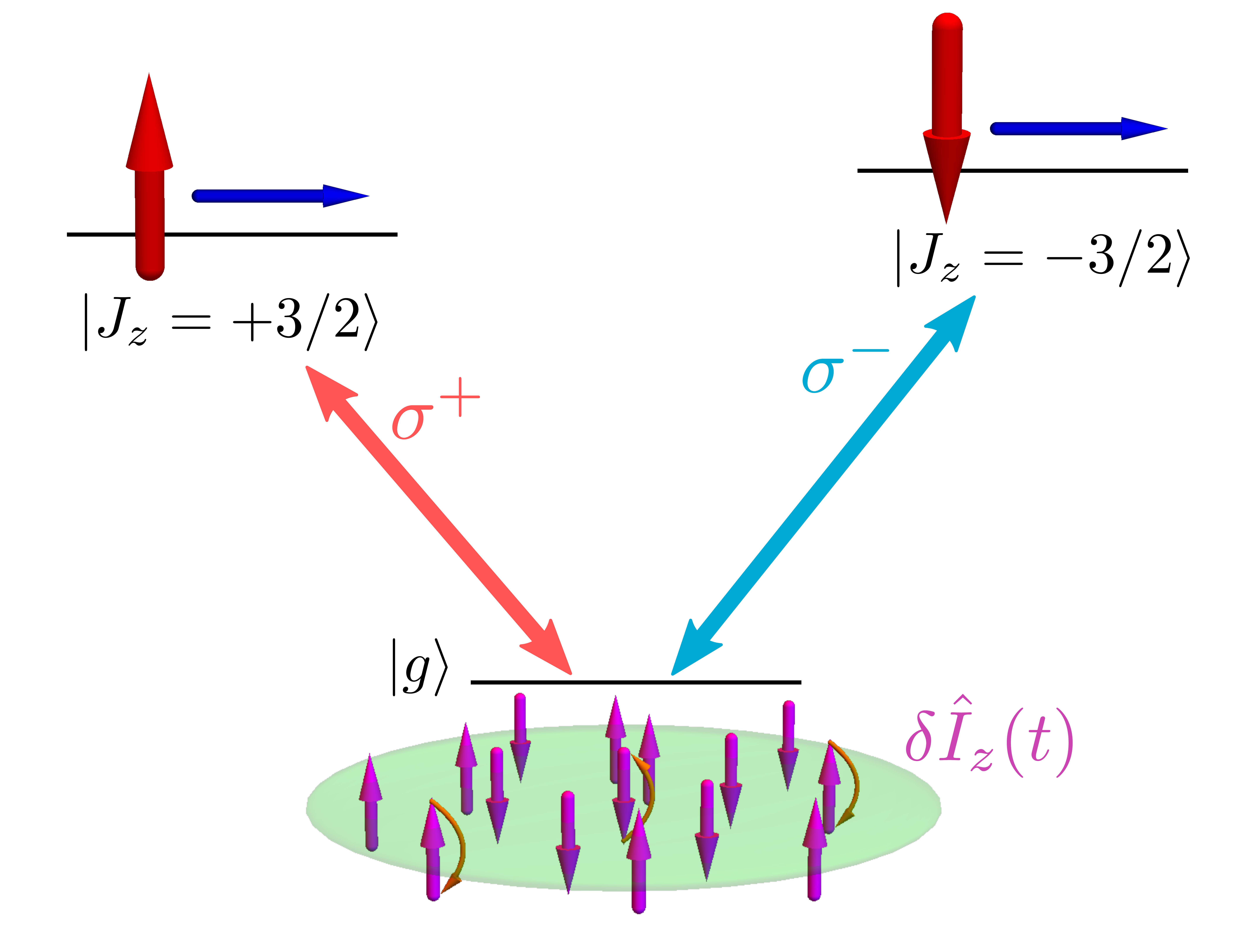}
  \caption{Quantum fluctuations of the {\Mn} spins (magenta arrows) in the exciton localization volume (green) randomly shift the exciton transitions energies. The states are characterized by the heavy hole spin (red arrow) $J_z=\pm3/2$, while the \addDima{projection of the electron spin (blue arrow) on the magnetic field does not change.} \commentDima{I updated the figure.}}
  \label{fig:V-scheme}
\end{figure}

We focus on the Voight geometry, when external magnetic field is applied along $x$ direction, perpendicular to the optical axis $z$. The {\Mn} spin Hamiltonian takes a form
\begin{equation}
  \label{eq:H0}
  \mathcal H_0=\hbar\Omega_LI_x,
\end{equation}
where
\begin{equation}
  \label{eq:I_tot}
  \bm I=\sum_{i=1}^N\bm I_i
\end{equation}
is the total spin of $N$ Mn atoms in the exciton localization volume and $\Omega_L=g\mu_B B/\hbar$ is the Larmor precession frequency in the magnetic field $B$ with $g$ being the $g$-factor and $\mu_B$ being the Bohr magneton. We assume the Mn concentration to be small enough to neglect their exchange interaction. This corresponds to the Mn concentration of a few percent or less.

We stress, that we consider here the optical transitions to the localized exciton state, while other types of transitions, e.g., to the trion or to the biexciton state can be described in a similar way. Let us make some other simplifying assumptions, which make the theory transparent and the results very illustrating. First, we neglect the transverse hole $g$-factor, which is usually very small~\cite{Mar99}. Second, we assume, that the \addDima{$x$ projection of the electron spin does not change.} This means, that the magnetic field is not weak, and the electron Zeeman energy exceeds the interaction strength with the random spin components $I_{i,y}$ and $I_{i,z}$. In fact, {\Mn} spins can be partially polarized along the magnetic field at low temperatures, and can create the effective exchange magnetic field along the same direction, which \addDima{can exceed the} external magnetic field for electrons. Below for simplicity we consider only one electron spin state (say, $S_x=+1/2$). This implicitly assumes, that the splitting of the electron spin sublevels exceeds the homogeneous and inhomogeneous widths of the optical resonance. Under these assumptions we arrive to the optical V-scheme, which is shown in Fig~\ref{fig:V-scheme}. Here the exciton vacuum state $\ket{g}$ can be excited by $\sigma^+$ of $\sigma^-$ polarized light to the exciton state with the heavy hole spins $J_z=\pm3/2$, respectively.

Without the exchange interaction with the magnetic impurities the two excitonic states are degenerate, so the exciton Hamiltonian reads
\begin{equation}
  \mathcal H_{exc}=\hbar\omega_0 n_{exc},
\end{equation}
where $\omega_0$ is the resonance frequency and $n_{exc}$ is the occupancy of the both exciton states. The exchange interaction leads to the splitting of the two resonances, which we describe by
\begin{equation}
  \mathcal H_{ex}=\hbar\omega_{ex}\frac{2}{3}S_z^hI_z.
\end{equation}
Here we neglect the total shift of the two resonances due to the exchange interaction with electron ($\omega_{ex,i}^e=0$) and consider the hole exchange interaction along the $z$ axis \addDima{only}~\cite{Merkulov_book}. Also for the simplicity we use the box model and set equal exchange interaction constants for all {\Mn} spins: $\omega_{ex,i}^{h,\alpha}=(2/3)\omega_{ex}\delta_{\alpha,z}$.
%In fact, the electron exchange interaction leads simply to the effective renormalization of Mn g-factor.

%Note, that the inhomogeneous broadening of the exciton resonance is usually much larger, than the homogeneous one, so the final results should be averaged over this frequency. 
Under these assumptions, the two excitonic states are not mixed. As a results, the circularly polarized $\sigma^\pm$ probe \addDima{light} induces the exciton dipole polarization with the same helicity $P_\pm(t)=[\mp P_x(t)-\i P_y(t)]/\sqrt{2}$. From Eq.~\eqref{eq:psi} we find, that
\begin{equation}
  \frac{\d P_{\pm}(t)}{\d t}=-\i\left[\omega_0\pm\omega_{ex}I_z(t)-\i\gamma\right]P_\pm(t)+\frac{\i}{\hbar}|d|^2E_\pm(t)\e^{-\i\omega_p t},
\label{eq:Pt}
\end{equation}
where $d$ is the optical transition dipole moment \addDima{%($|d|^2=P_\pm P_\pm^\dag$,
(see Appendix~\ref{app:P})}. This equation clearly shows, that the spin polarization $I_z(t)$ shifts the exciton resonance energy $\omega_0$, which allows one to optically monitor {\Mn} spin fluctuations. In fact, it follows from Eqs.~\eqref{eq:Hint_int} and~\eqref{eq:H0}, that $I_z(t)=\cos(\Omega_L t)I_z+\sin(\Omega_L t)I_y$, but we prefer to keep the general notation $I_z(t)$, which is valid for arbitrary spin Hamiltonian $\mathcal H_0$.

%For the electron spin parallel to the $x$ axes, the dipole moments of the two excitonic states in $\sigma^+$ and $\sigma^-$ polarizations are equal. While the generalization to the opposite case is trivial, 
In the specific system under study, the general expression~\eqref{eq:P_gen} for the polarization reduces to
  \begin{multline}
    \label{eq:P_main}
    P_\pm(t)=\i\frac{|d|^2}{\hbar}{E_\pm\e^{-\i\omega_p t}\int\limits_0^\infty\e^{\i(\omega_p-\omega_0)\tau-\gamma\tau}}\\
    \times\mathcal T\exp\left[\mp\i\omega_{ex}\int\limits_0^\tau I_z(t-\tau')\d\tau'\right]\d\tau.
  \end{multline}
Here the inner integral can be solved as described in Appendix~\ref{app:P}, but for the calculation of the spectra of the secondary emitted light this expression is more convenient.
%\commentDima{May be the outer integral can be solved as well?}

It is \addDima{useful} to analyze this expression in the adiabatic approximation. Provided $\Omega_L\ll\gamma$ the Mn spin dynamics is slow as compared \addDima{with} the exciton polarization relaxation. In this case one can replace $I_z(t-\tau')$ with $I_z(t)$ and solve the ordered exponential. Then the outer integral can be solved as well, which yields
\begin{equation}
  \label{eq:P_adiabat}
  P_\pm(t)=\frac{|d|^2E_\pm}{\hbar}\frac{\e^{-\i\omega_p t}}{\omega_0\pm\omega_{ex}I_z(t)-\omega_p-\i\gamma}.
\end{equation}
This expression shows that: (i) Mn spin polarization shifts the exciton resonance frequency, and (ii) the relation between Mn spin polarization and the exciton dipole polarization is nonlinear. The latter makes it possible to detect high order spin correlators using the resonance shift spin spectroscopy. Crucially, $I_z(t)$ should be considered here as the Heisenberg operator. Only when $\bm I$ is \addDima{a large classical vector, the corresponding} operator can be replaced with its expectation value.

\section{General relations between spin correlation functions and resonance shift optical response}
\label{sec:general}

The spin noise spectra are typically measured using \addDima{the} linearly polarized light,
  \begin{equation}
    \label{eq:Ex}
    \bm E=E_0\bm e_x,
  \end{equation}
where $E_0$ is an amplitude of the probe light, and $\bm e_x$ is a unit vector along $x$ axis.
%The Faraday rotation angle of the polarization plane, $\theta_F$, and the induced ellipticity of light, $\theta_E$,
The optical Faraday and ellipticity signals, $\mathcal F$ and $\mathcal E$, respectively, are measured in the transmission or reflection \addDima{geometry. They are given by the real and imaginary parts of the equality}~\cite{glazov:review}
\begin{equation}
  \mathcal F-\i\mathcal E=E_x'^*E_y'.
\end{equation}
Here $\bm E'$ is the amplitude of the emitted (or scattered) light. It consists of the contribution from the elastic scattering and \addDima{the} secondary emission by the exciton dipole polarization:
\begin{equation}
  \label{eq:E_prime}
  \bm E'=a \bm E + b \bm P,
\end{equation}
where $a$ and $b$ are complex coefficients, which depend on the geometry of the structure~\cite{book_Glazov}. In the next section we calculate the Faraday rotation noise spectra, which can be detected using the resonance shift spin spectroscopy.

The Faraday rotation angle of the probe polarization plane and the ellipticity angle can be calculated as
\begin{equation}
  \theta_F-\i\theta_E=\frac{\mathcal F-\i\mathcal E}{\mathcal I},
\end{equation}
where $\mathcal I=\left|E_x'\right|^2$ is proportional to the intensity of the detected light. These expressions assume, that the angles are small $\theta_{F,E}\ll 1$ or equivalently $|E_y'|\ll|E_x'|$.
% \begin{equation}
%   \theta_F+\i\theta_E=\frac{E_y'}{E_x'},
% \end{equation}
% where it is assumed that $\theta_{F,E}\ll1$.
Usually the scattering is weak, $bP_x\ll a E_0$, so by virtue of Eq.~\eqref{eq:E_prime} we arrive to
\begin{equation}
  \label{eq:thetas}
  \theta_F+\i\theta_E=\frac{E_y'}{E_x'}=\frac{b}{aE_0}P_y.
\end{equation}
This expression shows, that the spin signals are proportional to the exciton polarization along $y$ axis.

The optical signals noise spectra are defined as the Fourier transform of the \addDima{correlation} functions:
\begin{equation}
  \label{eq:spin_spec}
  (\theta_{F,E}^2)_\Omega=\int\limits_{-\infty}^\infty\braket{\theta_{F,E}(t)\theta_{F,E}(t+\tau)}_s\e^{\i\Omega\tau}\d\tau.
\end{equation}
Here the angular brackets denote quantum mechanical averaging and the subscript ``$s$'' denotes the symmetrized correlation function
\begin{equation}
  \braket{\theta(0)\theta(\tau)}_s=\frac{\braket{\theta(0)\theta(\tau)+\theta(\tau)\theta(0)}}{2}.
\end{equation}
In the steady state, the averages do not depend on time $t$, so for the rest of the paper we set $t=0$ in the correlation functions. The symmetrization is related with the fact, that the detected light is almost classical and the optical spin signals are self-homodyned~\cite{ll3_eng,Clerk,BackAction}. From Eqs.~\eqref{eq:spin_spec} and~\eqref{eq:thetas} one can see, that that the Faraday and ellipticity noise spectra are \addDima{determined} by the Fourier transform of the correlation functions of the components of the exciton polarization. \addDima{In Pauli-blocking spin noise spectroscopy this correlator is simply proportional to the spin noise spectrum.
%   \begin{equation}
%     \label{eq:spec_Iz}
%     (I_z^2)_\Omega=\int\limits_{-\infty}^\infty\braket{I_z(0)I_z(\tau)}\e^{\i\Omega\tau}\d\tau
%   \end{equation}
% (which should be symmetrized for low temperatures).
However, in the case of the resonance shift spin noise spectroscopy, there is no direct relation between them.}

In the same time the Raman spectrum of the scattered light in polarization $\alpha$ is given by
\begin{equation}
  S_{\rm tot}(\omega)=\int\limits_{-\infty}^\infty\braket{E_\alpha'^*(t)E_\alpha'(t+\tau)}\e^{\i\omega\tau}\d\tau.
\end{equation}
In the general case the Raman spectrum consists of a $\delta$-peak at $\omega=\omega_p$, which does not carry information about the spin system, and the rest of the spectrum $S(\omega)$. It can be presented as
\begin{equation}
  \label{eq:Raman}
  S(\Omega+\omega_p)=\left|b\right|^2\int\limits_{-\infty}^\infty\braket{P_\alpha^*(0)P_\alpha(\tau)}\e^{\i\Omega\tau}\d\tau.
\end{equation}
Thus we arrive again \addDima{at} the Fourier transform of the exciton polarization correlation function.

The exciton polarization is given by Eq.~\eqref{eq:P_main}, which implicitly depends on {\Mn} spin fluctuations. Now let us establish the general relations between the spectra of different polarization components and the multi-order spin correlation functions.
% on one hand, and \addDima{Faraday rotation and ellipticity noise spectra% as well as the Raman spectra
% } on the other hand. \addDima{We start the analysis by calculation of the Raman spin flip spectra, which then can be directly related to the noise spectra of Faraday and ellipticity spin signals.}

\subsection{Spectrum in circular polarization}
\label{sec:circ}

In this section we consider an auxiliary problem of $\sigma^+$ incident light, so the scattered light has the same polarization, $\alpha=+$. We calculate the spectra of the dimensionless exciton polarization
\begin{equation}
  \bm p(t)=-\frac{\hbar\gamma}{|d|^2E_0}\bm P(t)\e^{\i\omega_p t},
\end{equation}
which in this case are proportional to the Raman spin flip spectrum in $\sigma^+$ polarization, see Eq.~\eqref{eq:Raman}. The spectrum in $\sigma^-$ polarization is the same.

It is convenient to rewrite Eq.~\eqref{eq:P_main} as
  \begin{equation}
    \label{eq:p}
    p_+(t)=\int\limits_0^\infty\e^{-\i\delta k-k}\mathcal T\exp\left[-\i\mathcal J(t)\right]\d k,
  \end{equation}
where
\begin{equation}
  \delta=(\omega_0-\omega_p)/\gamma,
\end{equation}
is a dimensionless detuning and
\begin{equation}
  \label{eq:J_def}
  \mathcal J(t)=\int\limits_0^{k}m(t-k'/\gamma)\d k'
\end{equation}
with
\begin{equation}
  \label{eq:m_def}
  m(t)=\frac{\omega_{ex}}{\gamma}I_z(t)
\end{equation}
being a dimensionless splitting of the resonance. Further, we note that for the localized excitons, the inhomogeneous broadening usually exceeds by far the homogeneous one (see, e.g., Ref.~\onlinecite{Stuhler1995}). Therefore the spin noise and Raman spin flip spectra should be averaged over the detuning as
\begin{equation}
  \label{eq:aver_def}
  \overline{\braket{p_+^*(0)p_+(\tau)}}=\frac{1}{2\pi}\int\limits_{-\infty}^\infty\braket{p_+^*(0)p_+(\tau)}\d\delta.
\end{equation}
Here we introduced the factor $1/(2\pi)$ to shorten the following expressions. We substitute \addDima{here} the exciton polarization from Eq.~\eqref{eq:p} and obtain the averaged correlation function
\begin{equation}
  \label{eq:pp}
  \overline{\braket{p_+^*(0)p_+(\tau)}}
  =\int\limits_0^\infty\e^{-2k}\braket{\left[\overline{\mathcal T}\e^{\i\mathcal J(0)}\right]\left[\mathcal T\e^{-\i\mathcal J(\tau)}\right]}\d k,
\end{equation}
where $\overline{\mathcal T}$ denotes the reversed time ordering. The correlator in this expression can be calculated using the cumulant expansion.

Generally, the quantum noise statistics is completely described by the series of cumulants of the random variable~\cite{Kubo_Cumulant,Mendel,Gardiner}. \addDima{In Appendix~\ref{app:cumulants} we obtain the general expressions for the polarization correlation function and simplify it for the Gaussian spin noise.}
To obtain a \addDima{simple expression} for the polarization correlator let us consider again \addDima{the} adiabatic approximation, $\Omega_L\ll\gamma$. In this case one can \addDima{replace $m(t-k'/\gamma)$ in Eq.~\eqref{eq:J_def} with $m(t)$. % calculate $\braket{\mathcal J(0)\mathcal J(\tau)}$ and $\braket{\mathcal J^2(0)}_s$. Using
Then from Eq.~\eqref{eq:pp} we obtain
\begin{equation}
  \label{eq:pp_int_one}
  \overline{\braket{p_+^*(0)p_+(\tau)}}=\sum_{n=0}^\infty\sum_{l=0}^{2n}\frac{(-1)^{n+l}}{2^{2n+1}}{{2n}\choose{l}}\braket{m^l(0)m^{2n-l}(\tau)}\addDima{,}
\end{equation}
\addDima{where ${{2n}\choose{l}}$ is the binomial coefficient.}
This expression shows, that the polarization correlator and Faraday rotation noise spectra are determined by the spin correlation functions of all orders. Thus, the resonance shift spin noise spectroscopy measures high order spin correlators in addition to the standard second order correlator. This surprising result originates in the nonlinear relation between the exciton polarization and the total {\Mn} spin in the limit $I_z\gg 1$ [see, e.g., Eq.~\eqref{eq:P_adiabat}].}

\addDima{For Gaussian spin noise using Eq.~\eqref{eq:pp_int_J} we obtain}
\begin{equation}
  \label{eq:pp_int_one}
  \overline{\braket{p_+^*(0)p_+(\tau)}}=\int\limits_0^\infty\e^{-k^2\Delta m^2(\tau)}\e^{-2k}\d k,
\end{equation}
where we introduced
\begin{equation}
  \Delta m^2(\tau)=\braket{m^2}-\braket{m(0)m(\tau)}.
\end{equation}
This integral can be solved as
\begin{multline}
  \label{eq:Pp_int}
  \overline{\braket{p_+^*(0)p_+(\tau)}}=\\
  \frac{1}{2}\sqrt{\frac{\pi}{\Delta m^2(\tau)}}\e^{1/\Delta m^2(\tau)}\erfc\left(\frac{1}{\sqrt{\Delta m^2(\tau)}}\right).
\end{multline}

If exchange interaction is weak, $m(t)\ll1$ one can use the asymptotic expansion
\begin{equation}
  \label{eq:m0}
  \overline{\braket{p_+^*(0)p_+(\tau)}}=\frac{1}{2}\sum_{n=0}^\infty\left[-\frac{\Delta m^2(\tau)}{2}\right]^n(2n-1)!!\:.
\end{equation}
This expression directly relates the spin correlation function $\Delta m^2(\tau)$ with the polarization correlator.

\subsection{Faraday rotation noise spectra\\ and Raman spectra in linear polarizations}
\label{sec:linear}

Let us return to the linearly polarized probe light, Eq.~\eqref{eq:Ex}. Similarly to Eq.~\eqref{eq:p}, the dimensionless exciton polarization in this case reads
\begin{subequations}
  \label{eq:Py_f}
  \begin{equation}
    p_x(t)=\int\limits_{0}^\infty\e^{-\i\delta k-k}\mathcal T\cos\left[\mathcal J(t)\right]\d k,
  \end{equation}
  \begin{equation}
    \label{eq:Py_f_2}
    p_y(t)=-\int\limits_{0}^\infty\e^{-\i\delta k-k}\mathcal T\sin\left[\mathcal J(t)\right]\d k,
  \end{equation}
\end{subequations}
% where $\mathcal T\sin(x)$ is defined below Eq.~\eqref{eq:Py_gen} and $\mathcal T\cos(x)$ is defined in an analogous way. 
where we introduced the notations
\begin{equation}
  \mathcal T\cos(x)=\frac{\mathcal T\e^{\i x}+\mathcal T\e^{-\i x}}{2},
  \quad
  \mathcal T\sin(x)=\frac{\mathcal T\e^{\i x}-\mathcal T\e^{-\i x}}{2\i}.
\end{equation}
Then one can perform the calculations following the lines of the previous subsection: average the correlation function $\braket{p_\alpha^*(0)p_\alpha(\tau)}$ over the detuning, express it through the cumulants and in the limit $N\gg1$ neglect all the cumulants of high orders. For clarity we give the final result for the adiabatic limit [c.f. \addDima{Eq.~\eqref{eq:pp_int_one}}]:
% Then one can perform the calculations \addDima{following the lines of} the previous subsection: average the correlation function $\braket{p_\alpha^*(0)p_\alpha(\tau)}$ over the detuning as in Eq.~\eqref{eq:aver_def}, express it through the cumulants as in Eq.~\eqref{eq:JJ_gen} and under the assumption of $N\gg1$ neglect all the cumulants of high orders as in Eq.~\eqref{eq:pp_int_J}. For clarity we give the final result for the adiabatic limit (c.f. Eq.~\eqref{eq:pp_int_J}):
\begin{widetext}
  \begin{equation}
   \label{eq:Py_int}
    \begin{bmatrix} \overline{\braket{p_x^*(0)p_x(\tau)}} \\[2mm] \overline{\braket{p_y^*(0)p_y(\tau)}} \end{bmatrix}
    =\int\limits_0^\infty\d k\e^{-2k-k^2\braket{m^2}}
    \begin{bmatrix} \cosh\left(k^2\braket{m(0)m(\tau)}\right) \\[2mm] \sinh\left(k^2\braket{m(0)m(\tau)}\right) \end{bmatrix}
    .
  \end{equation}
\end{widetext}
These expressions along with Eq.~\eqref{eq:Raman} allow one to directly calculate the Raman spectrum \addDima{for the given Mn spin correlation function}. Here the integrals can be expressed through the error function, and its asymptotic expansions can be found. However, these expressions are cumbersome, and will not be needed below.

As a first step towards the relation between Raman and spin noise noise spectrum, we consider the Raman spectrum in crossed ($y$) polarization in the limit $\braket{m(t)m(t+\tau)}\ll1$. In this case the relation between the exciton dipole polarization and Mn spin polarization it linear, so from Eq.~\eqref{eq:Py_int} we obtain the spectrum% has~\cite{Gorbovitskii}
\begin{equation}
  S(\omega)=\left|b\frac{|d|^2E_0\omega_{ex}}{\hbar\gamma^2}\right|^2(I_z^2)_{\omega-\omega_p},
\end{equation}
where the spin noise noise spectrum of Mn$^{2+}$ ions
% fluctuations spectrum of localized spins (e.g those of Mn$^{2+}$ ions)
is given by
\begin{equation}
  \label{eq:spec_Iz} 
  (I_z^2)_{\Omega}=\int\limits_{-\infty}^\infty\braket{I_z(0)I_z(\tau)}\e^{\i\Omega\tau}\d\tau.
\end{equation}
%according to the quantum Wiener-Khinchin theorem.
An analogous relation between the Raman spin flip and the spin noise spectra was derived by Gorbovitskii and Perel for Pauli-blocking spin \addDima{noise} spectroscopy~\cite{Gorbovitskii}. Below we \addDima{demonstrate}, that a similar relation holds between multispin flip spectrum and the noise spectrum of optical signals for any strength of the exchange interaction.

Indeed, from Eq.~\eqref{eq:thetas} we obtain the correlation functions of the Faraday rotation and ellipticity angles
\begin{subequations}
  \label{eq:Py_corr}
  \begin{equation}
    \braket{\theta_F(0)\theta_F(\tau)}=\left|\frac{|d|^2b}{\hbar\gamma a}\right|^2\braket{p_y''(0)p_y''(\tau)}_s,
  \end{equation}
  \begin{equation}
    \braket{\theta_E(0)\theta_E(\tau)}=\left|\frac{|d|^2b}{\hbar\gamma a}\right|^2\braket{p_y'(0)p_y'(\tau)}_s,
  \end{equation}
\end{subequations}
where one and two primes denote the real and imaginary parts of the polarization, respectively. Similarly to Eq.~\eqref{eq:pp}, we average the correlation functions of $p_y'(\tau)$ and $p_y''(\tau)$ over the detuning \addDima{making use of} Eq.~\eqref{eq:Py_f_2}, and obtain
\begin{equation}
  \label{eq:pys}
  \overline{\braket{p_y''(0)p_y''(\tau)}}_s=\overline{\braket{p_y'(0)p_y'(\tau)}}_s=\frac{1}{2}\overline{\braket{p_y^*(0)p_y(\tau)}}_s.
\end{equation}
This expression differs from the \addDima{second} line of Eq.~\eqref{eq:Py_int} by a factor and symmetrization. Thus the noise \addDima{spectra} of the Faraday rotation and ellipticity angles %[see Eq.~\eqref{eq:thetas}] 
can be calculated as a symmetrized Raman spectrum in crossed linear polarizations. % [see Eq.~\eqref{eq:Raman}].
In the next section we calculate and describe these spectra.

%====================================================
\section{Faraday rotation noise spectrum in Voigt geometry}
\label{sec:indirect_spectra}

\subsection{Spin correlation functions}
\label{sec:spectra}

The general expressions~\eqref{eq:Py_int} relate the Faraday rotation noise and Raman spin flip spectra with the spin correlators. Here we calculate the correlation functions in external magnetic field described by Eq.~\eqref{eq:H0}.

The average Mn spin is oriented along $\bm B$ and equals to
\begin{equation}
  \braket{I_x}=Ns\mathcal B_{s}\left(\frac{g\mu_{B}Bs}{k_B T}\right),
\end{equation}
where $s=5/2$ is a single {\Mn} spin, $\mathcal B_{s}(x)$ is the Brillouin function, $k_B$ is the Boltzmann constant, and $T$ is the temperature. Using the commutation relations for the spin components we find also the correlators
\begin{subequations}
  \label{eq:I2_init}
  \begin{equation}
    \braket{I_yI_z}=-\braket{I_zI_y}=\frac{\i}{2}\braket{I_x},
  \end{equation}
  \begin{equation}
    \label{eq:Iz2_init}
    \braket{I_z^2}=\braket{I_y^2}=\frac{N}{2}\left[s(s+1)-\braket{s_x^2}\right].
  \end{equation}
\end{subequations}
Notably, the first of these two equations is responsible for the quantum part of the spin correlation functions. Indeed, for classical noise the correlation functions do not depend on the order, in which the fluctuating quantities are multiplied. Moreover, Eq.~\eqref{eq:Iz2_init} shows, that even at zero temperature, when $\braket{s_x^2}=s^2$, zero-point spin fluctuations $\braket{I_z^2}=Ns/2$ are present.

% The limit of low spin temperature is realized in the magnetic fields of the order of several Tesla at liquid helium temperatures. Experimentally detection of the quantum spin noise spectra in the corresponding frequency can be challenging and require a pulse trains~\cite{Berski-fast-SNS} or heterodyne detection~\cite{Heterodyne}.

% The zero spin fluctuations become important not only when the temperature is low, but also when the magnetic field is strong enough, to have macroscopic spin polarization $\braket{I_x}$. Thus The observation of zero-point \Mn spin fluctuations can be done at temperatues $T\sim4$~K in magnetic fields $B\gtrsim3$~T.

The time correlation functions for $\tau>0$ obey the equations
\begin{subequations}
  \label{eq:dyn}
  \begin{equation}
    \frac{\d}{\d\tau}\braket{I_z(0)I_z(\tau)}=\Omega_L\braket{I_z(0)I_y(\tau)}-\frac{\braket{I_z(0)I_z(\tau)}}{\tau_s},
  \end{equation}
  \begin{equation}
    \frac{\d}{\d\tau}\braket{I_z(0)I_y(\tau)}=-\Omega_L\braket{I_z(0)I_z(\tau)}-\frac{\braket{I_z(0)I_y(\tau)}}{\tau_s},
  \end{equation}
\end{subequations}
where $\tau_s$ is a transverse spin relaxation time \addDima{(we assume that $\hbar/\tau_s\ll k_BT$)}. The solution of these equations with the initial conditions~\eqref{eq:I2_init} reads
  \begin{multline}
    \label{eq:Iz2}
    \braket{I_z(0)I_z(\tau)}=\frac{1}{2}\left[\left(\braket{I_z^2}+\frac{\braket{I_x}}{2}\right)\e^{\i\Omega_L\tau}\right.\\
    +\left.\left(\braket{I_z^2}-\frac{\braket{I_x}}{2}\right)\e^{-\i\Omega_L\tau} \right]\e^{-|\tau|/\tau_s}.
  \end{multline}
Then using the definition~\eqref{eq:m_def} we find the dimensionless \addDima{correlation} function
\begin{equation}
  \label{eq:m2}
  \braket{m(0)m(t)}=\left(\mu_+\e^{-\i\Omega_L\tau}+\mu_-\e^{\i\Omega_L\tau}\right)\e^{-|\tau|/\tau_s},
\end{equation}
where we introduced
\begin{equation}
  \label{eq:mu_pm}
  \mu_\pm=\frac{\omega_{ex}^2}{2\gamma^2}\left(\braket{I_{z}^2}\mp\frac{\braket{I_{x}}}{2}\right).
\end{equation}
The correlator $\braket{m(0)m(t)}$ ultimately defines the noise spectra of Faraday rotation and ellipticity, which we analyze in the next subsection.

\subsection{Faraday rotation noise and Raman spectra}
% \subsection{Calculation of noise spectra of spin signals}

Similarly to Sec.~\ref{sec:general} it is convenient to start from the analysis of the exciton dipole polarization \addDima{correlation function} in circular polarizations, which define the corresponding Raman spectra.

\addDima{
To shorten the notation we introduce the scaled spectrum [c.f. Eq.~\eqref{eq:Raman}]
\begin{equation}
  \label{eq:S_tilde}
  \mathcal S_{++}(\Omega)=\int\limits_{-\infty}^\infty\overline{\braket{p_+^*(0)p_+(\tau)}}\e^{\i\Omega\tau}\d\tau.
\end{equation}
From Eqs.~\eqref{eq:m0} and~\eqref{eq:m2} one can see, that the spectrum consists of the peaks at frequencies $n\Omega_L$, where $n$ is an integer. The general form of the spectrum of circularly polarized exciton dipole polarization is
\begin{equation}
  \label{eq:spec_all}
  \mathcal S_{++}(\Omega)=\sum_{n=1}^\infty\sum_\pm \mathcal P_n^\pm(\Omega\mp n\Omega_L),
\end{equation}
where $P_n^\pm(\Omega)$ are even functions peaked at zero and we neglect the peak at zero frequency.}

%\commentDima{Boltzmann ratio and $\mu_\pm$.}
\addDima{Another general property follows from the ratio of prefactors in Eq.~\eqref{eq:m2}. From the definition~\eqref{eq:spec_Iz} (without symmetrization) one can see, that
  \begin{equation}
    \frac{(I_z^2)_{\Omega_L}}{(I_z^2)_{-\Omega_L}}=\e^{-\hbar\Omega_L/(k_B T)}.
  \end{equation}
Generally, this relation is inherited by the polarization spectra [Eq.~\eqref{eq:S_tilde}] in the form
\begin{equation}
  \frac{\mathcal S_{++}(\Omega)}{\mathcal S_{++}(-\Omega)}=\e^{-\hbar\Omega/(k_B T)},
\end{equation}
which is well known for the Raman spectra.}

\addDima{
The Raman spin flip spectra in $\sigma^+$ and in $\sigma^-$ polarizations coincide and are given by Eq.~\eqref{eq:spec_all}:
\begin{equation}
  \mathcal S_{--}(\Omega)=\mathcal S_{++}(\Omega).
\end{equation}
We recall, that the two circular polarizations are independent in the lowest order in the incident electric field, so the cross-polarized spectra in circular polarizations vanish: $\mathcal S_{\pm\mp}(\Omega)=0$.
}

\addDima{
The Raman spin flip spectra in linear polarizations and Faraday rotation and ellipticity noise spectra can be calculated using Eqs.~\eqref{eq:Py_int} and~\eqref{eq:pys} for the adiabatic regime. Similarly to Eq.~\eqref{eq:spec_all} the Raman spectra take the form
  \begin{subequations}
    \label{eq:S_lin}
    \begin{equation}
      \mathcal S_{xx}(\Omega)=\mathcal S_{yy}(\Omega)=\sum_{k=1}^\infty\sum_\pm\mathcal P_{2k}^\pm(\Omega\mp 2k\Omega_L),
    \end{equation}
    \begin{equation}
      \mathcal S_{xy}(\Omega)=\mathcal S_{yx}(\Omega)=\sum_{k=0}^\infty\sum_\pm\mathcal P_{2k+1}^{\pm}(\Omega\mp(2k+1)\Omega_L).
    \end{equation}
  \end{subequations}
  The Faraday rotation (and ellipticity) noise spectra are given by
  \begin{equation}
    \label{eq:S_sn}
    \mathcal S_{FR}(\Omega)=\frac{\mathcal S_{xy}(\Omega)+\mathcal S_{xy}(-\Omega)}{2},
  \end{equation}
%\commentDima{May be it is better to replace $\mathcal S_{sn}(\Omega)$ with $\mathcal S_{FR}(\Omega)$?}
which differs from $\overline{(\theta_F^2)_\Omega}$ (and $\overline{(\theta_E^2)_\Omega}$) by a factor. %Thus, we obtain the general expressions~\eqref{eq:An_pm} and~\eqref{eq:Pn} for the amplitudes and the width of the peaks in the spin noise and Raman spectra, respectively.
}

\addDima{In the illustrative case} of $m(t)\ll1$ one can substitute Eq.~\eqref{eq:Iz2} in the asymptotic Eq.~\eqref{eq:m0}, which \addDima{yields
\begin{equation}
  \label{eq:S_m_small}
  \mathcal S_{++}(\Omega)=\sum_{n=1}^\infty\sum_\pm\frac{(2n-1)!!}{2^{n}}\mu_\pm^n\frac{\tau_s/n}{1+\left[(\Omega/n\mp\Omega_L)\tau_s\right]^2}\addDima{.}
\end{equation}
%where we neglected the peak at zero frequency.
%  and introduced the Lorentzian functions
% \begin{equation}
%   \label{eq:Pn}
%   \mathcal P_n(\Omega)=\frac{2\tau_s/n}{1+\left[(\Omega/n-\Omega_L)\tau_s\right]^2}.
% \end{equation}
One can see, that the spectrum consists of Lorentzian} peaks at the frequencies $\mp n\Omega_L$ with the areas (divided by $2\pi$)
\begin{equation}
  \label{eq:A_n_small}
  A_n^\pm=\frac{(2n-1)!!}{2^{n+1}}\mu_\pm^n,
\end{equation}
respectively. The widths of these peaks are $n/\tau_s$. We stress, that the second order spin correlation function~\eqref{eq:Iz2} contains the spin precession frequencies $\pm\Omega_L$ only. Therefore the appearance of the overtones is a \addDima{fingerprint} of the contributions of the higher order spin correlators.

For high order contributions, $n\gg 1$, Eq.~\eqref{eq:S_m_small} diverges, and the above analysis is inapplicable. Generally, one has to start from Eq.~\eqref{eq:pp_int_one}, where
\begin{equation}
  \Delta m^2(\tau)=\sum_\pm\mu_\pm\left(1-\e^{\mp\i\Omega_L\tau}\e^{-|\tau|/\tau_s}\right)
\end{equation}
according to Eq.~\eqref{eq:m2}. Thus we obtain
\begin{multline}
  \overline{\braket{p_+^*(0)p_+(\tau)}}=\int\limits_0^\infty\e^{-2k}\e^{-k^2(\mu_++\mu_-)}\\
  \times\exp\left[k^2\left(\mu_+\e^{-\i\Omega_L\tau}+\mu_-\e^{\i\Omega_L\tau}\right)\e^{-|\tau|/\tau_s}\right]\d k.
\end{multline}
The exponent in the second line can be decomposed in the Taylor series as
\begin{multline}
  \label{eq:corr_power}
  \overline{\braket{p_+^*(0)p_+(\tau)}}=\int\limits_0^\infty\e^{-2k}\e^{-k^2(\mu_++\mu_-)}\\
  \times\sum_{n=0}^{\infty}\frac{k^{2n}}{n!}\left(\mu_+\e^{-\i\Omega_L\tau}+\mu_-\e^{\i\Omega_L\tau}\right)^n\e^{-n|\tau|/\tau_s}\d k.
\end{multline}
Again, one can see, that the correlation function contains harmonics $\propto\addDima{\e^{\mp\i n\Omega_L\tau}}$, so the spectrum consists of the peaks at frequencies $\pm n\Omega_L$.

\begin{figure}
  \includegraphics[width=\linewidth]{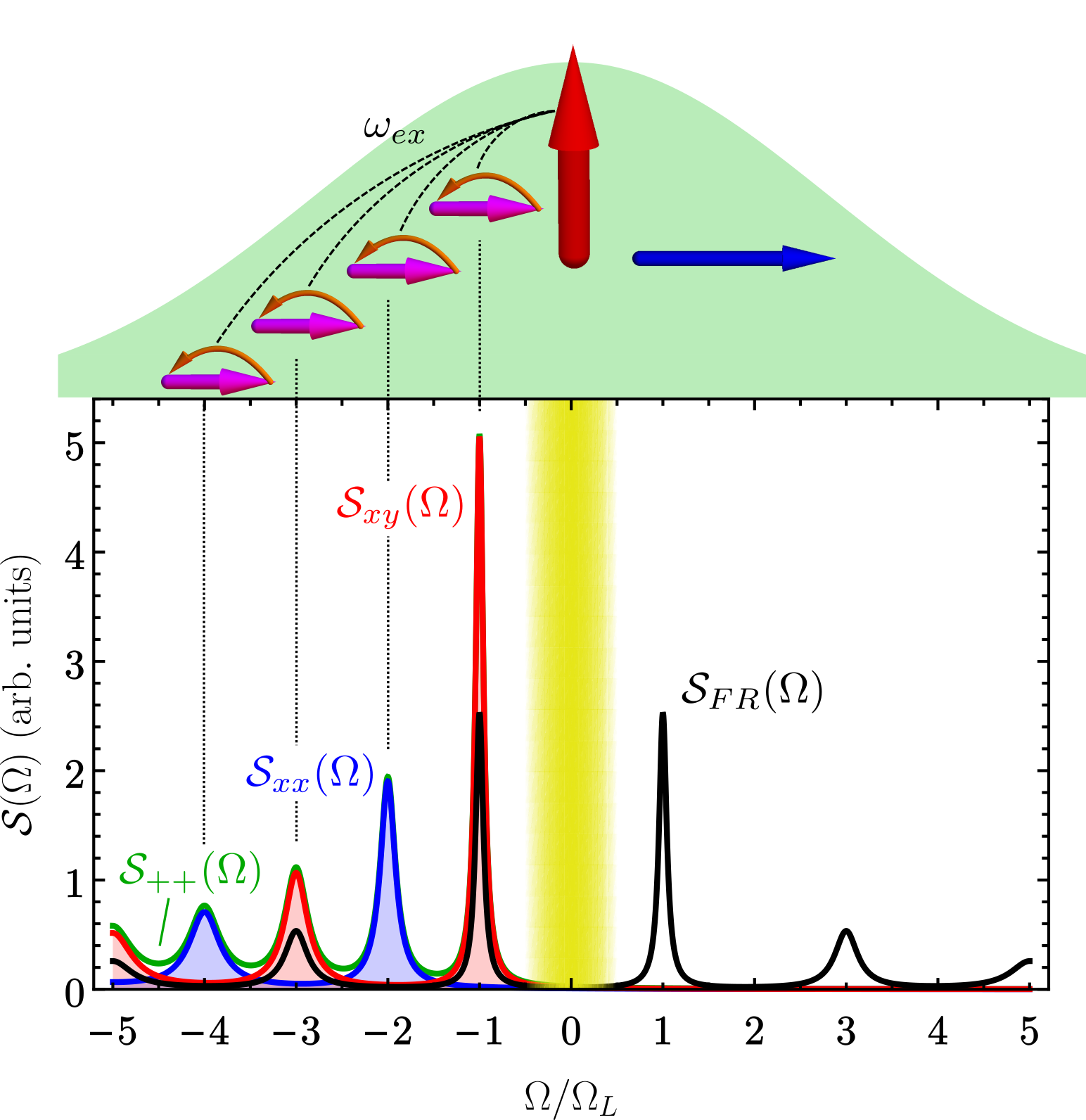}
  \caption{The noise spectrum of Faraday rotation, \addDima{$\mathcal S_{FR}(\Omega)$}, \commentDima{I changed this in the figure.} (black line), and multispin flip Raman spectra $\mathcal S_{xx}(\Omega)$ (blue line with filling), $\mathcal S_{xy}(\Omega)$ (red line with filling), and \addDima{$\mathcal S_{++}(\Omega)$} (green line), calculated after Eqs.~\eqref{eq:S_sn}, \eqref{eq:S_lin}, and~\eqref{eq:spec_all}, respectively. The parameters of the calculation are $T=0$, $\sqrt{N}\omega_{ex}/\gamma\gg1$, and $\tau_s\Omega_L=20$. The peaks correspond to the multiple flips of {\Mn} spins (magenta arrows \addDima{in the sketch}) mediated by the exchange interaction with the heavy hole spin (red arrow) in the exciton localization area (green).
}
  \label{fig:compare}
\end{figure}

In Fig.~\ref{fig:compare} we compare the different spectra in the limit of zero temperature $T=0$ and strong exchange interaction $\sqrt{N}\omega_{ex}\gg\gamma$. All the spectra show the comb of peaks at frequencies, which are multiples of the Larmor spin precession frequency. The peaks in the Raman spectra correspond to the multiple spin flips mediated by the excitonic state, as shown in the top of the figure. Importantly, multiple spin flips take place in one process due to the RKKY-type exchange interaction between {\Mn} spins mediated by the heavy hole spin. The same mechanism can also lead to the double spin flips of donor bound electrons~\cite{PhysRevLett.29.107} and of electrons confined in the nanoplatelets~\cite{DSF_NPL}. In the Faraday rotation noise spectra, the peaks at frequencies $\pm n\Omega_L$ reflect the contributions of quantum spin noise correlation functions of the order $2n$. We recall, that the average spin polarization along $z$ axis is absent, so the correlators of the odd orders vanish. Observation of high order spin correlators is possible due to the nonlinear relation between dipole polarization and the total {\Mn} spin. Similarly, noise of the linear birefringence was recently shown to produce the peak at the double Larmor frequency for cesium atoms~\cite{fomin2019spinalignment}.

The Raman spin flip spectra in Fig.~\ref{fig:compare} are asymmetric (contain only the peaks at negative frequencies) because the energy can not be absorbed from the zero-point spin fluctuations. Alternatively one can say, that the {\Mn} spins are all oriented along $x$ axis and can be flipped only in the opposite direction. The Raman spectra in linear polarizations are similar, but co-polarized (blue curve) and cross-polarized (red curve) spectra consist of the peaks at even and odd frequencies only, respectively, see \addDima{Eqs.}~\eqref{eq:S_lin}. In the same time, the Faraday rotation noise spectrum (\addDima{black} curve) is symmetric and contains the odd peaks only, as it follows from Eq.~\eqref{eq:S_sn}.

In the limit of zero \addDima{temperature,} the expressions for the spectra are particularly simple even beyond the adiabatic approximation. From the spin correlation function~\eqref{eq:m2} and Eqs.~\eqref{eq:J2} we find that
  \begin{subequations}
    \begin{equation}
      \braket{\mathcal J(0)\mathcal J(\tau)}=\braket{\mathcal J^2(0)}_s\e^{\i\Omega_L\tau-|\tau|/\tau_s},
    \end{equation}
    \begin{equation}
      \braket{\mathcal J^2(0)}_s=Ns\frac{\omega_{ex}^2}{\Omega_L^2}\left[1-\cos(\Omega_L k/\gamma)\right],
      \end{equation}
    \end{subequations}
  where we took into account that $\tau_s\gamma\gg 1$. Substitution of these expressions in the polarization correlation function~\eqref{eq:pp_int_J} yields the areas of the peaks in the form~\cite{Stuhler1995,Stuhler1996,ClassicallyLarge}
  \begin{equation}
    \label{eq:A_Mer}
    A_n^-=\addDima{\frac{1}{2\tau_0}}\int\limits_0^\infty\e^{-t/\tau_0}\frac{\Delta I_x^n(t)}{n!}\e^{-\Delta I_x(t)}\d t.
  \end{equation}
  Here we introduced the notations $t=k/\gamma$, $\tau_0=1/(2\gamma)$ and~\footnote{The integral in Eq.~\eqref{eq:A_Mer} converges at $\Delta I_x(t)\lesssim1$. Since $N\gg 1$ one has either $\omega_{ex}\ll\Omega_L$ or $\Omega_L t\ll 1$, so one can replace $\Omega_{\rm tot}$ with $\Omega_L$ and find $\Delta I_x(t)=\left<\mathcal J^2(0)\right>_s$.}
  \begin{equation}
    \label{eq:Delta_Ix}
    \Delta I_x(t)=I\frac{\omega_{ex}^2}{\Omega_{\rm tot}^2}\left[1-\cos(\Omega_{\rm tot}t)\right]
  \end{equation}
  with $I=Ns$ and $\Omega_{\rm tot}=\sqrt{\omega_{ex}^2+\Omega_L^2}$. Physically, $\Delta I_x(t)$ is the change of $I_x$ during the \addDima{spin} precession in the sum of the exchange and external magnetic fields for the time $t$. The integrand in Eq.~\eqref{eq:A_Mer} has a \addDima{form} of the probability of the change of $I_x$ by $n$ in the Poisson distribution. The integration describes the average of this probability during the exponential exciton decay described by $\e^{-t/\tau_0}$.

% The inset in Fig.~\ref{fig:temperature} shows the peaks' areas for different temperatures. For high temperature (black dots) the areas are symmetric, $A_n^+=A_n^-$, and decay \addDima{faster than} exponentially:
% \begin{equation}
%   \label{eq:An_exp}
%   A_n^\pm\propto 1/(\sqrt{n}2^n),
% \end{equation}
% see Eq.~\eqref{eq:An_strong_large}. With decrease of the temperature the spectra become asymmetric: the peaks at positive frequencies decay faster, and at negative frequencies --- slower.
%\addDima{In the limit of zero temperature, the peaks at positive frequencies disappear completely, but at negative frequencies decay} obeying the slow power law $1/\sqrt{n}$.% Surprisingly, Eq.~\eqref{eq:An_strong_large} describes reasonably well the peaks' areas for any $n$, as shown by the solid lines.

%\commentDima{May be add a description of the areas of the peaks?}

\begin{figure}
  \includegraphics[width=\linewidth]{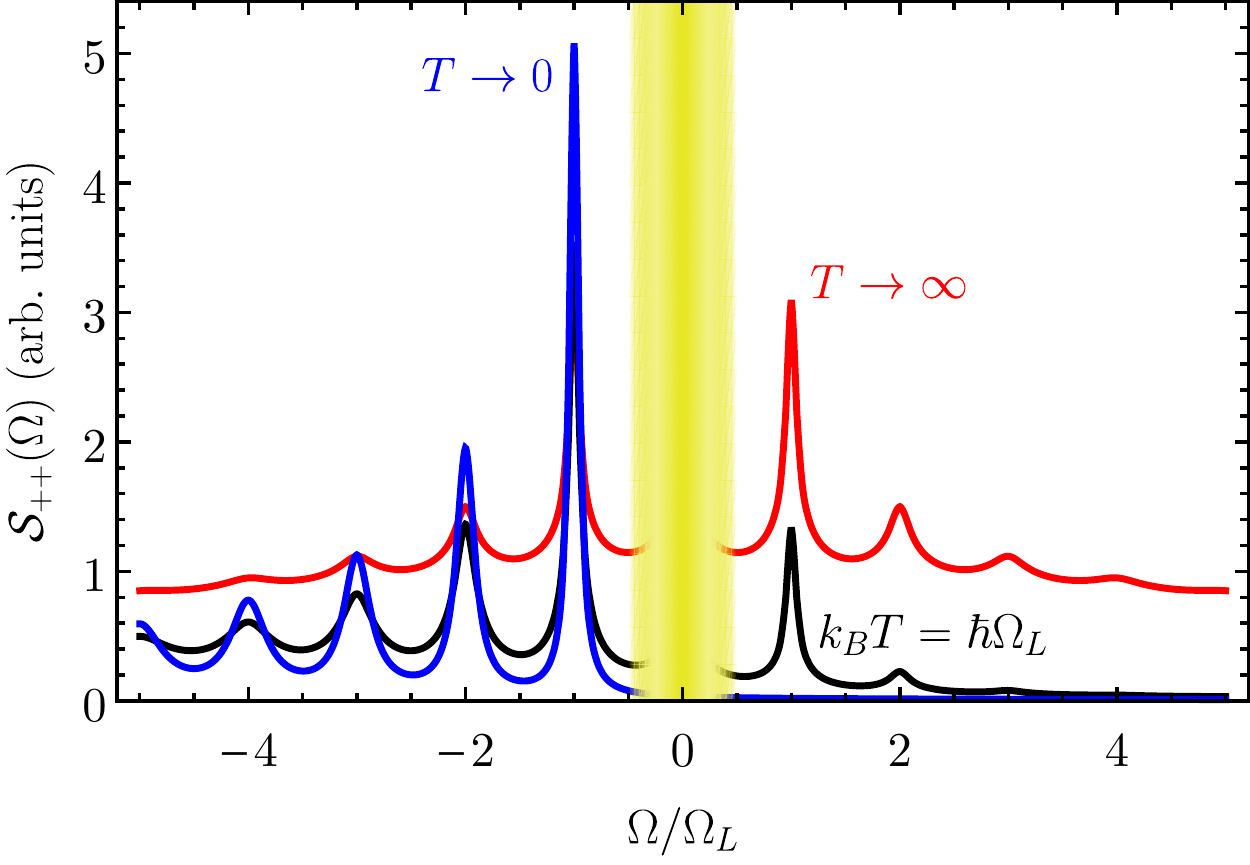}
  \caption{\addDima{Circular dipole polarization noise spectra, $\mathcal S_{++}(\Omega)$, calculated after Eq.~\eqref{eq:spec_all} in the limit of strong exchange interaction ($\sqrt{N}\omega_{ex}/\gamma\gg1$) for $\tau_s\Omega_L=20$ and for different temperatures, as indicated in the plot.}
% The inset shows with the same colors the areas of the peaks. The dots  and the curves are calculated after exact Eq.~\eqref{eq:An_strong} and approximate Eq.~\eqref{eq:An_strong_large}, respectively.
}
  \label{fig:temperature}
\end{figure}

Fig.~\ref{fig:temperature} shows the spectra for different temperatures, or equivalently for different magnetic fields. For better visibility we focus on the Raman spectrum $\addDima{\mathcal S}_{++}(\Omega)$, while the Faraday rotation noise spectrum can be obtained by selecting the odd numbered peaks and symmetrizing them in frequency, see Eq.~\eqref{eq:S_sn}. At high temperature (red curve), the spectrum is symmetric, which corresponds to purely thermal spin fluctuations. \addDima{In the limit $\omega_{ex}\sqrt{N}\gg\gamma$, the peaks are very broad and strongly overlap. At high frequencies the spectrum is described by
  \begin{equation}
    \mathcal S_{++}(\Omega)=\frac{\pi\gamma}{\omega_{ex}}\sqrt{\frac{3\tau_s}{2Ns(s+1)|\Omega|}}.
  \end{equation}
The fact that $\mathcal S_{++}(\Omega)$ decreases with increase of $\omega_{ex}$ is caused by the large splitting of the two excitonic resonances in this limit and a small region of values of $\delta I_z$, which produces sizable Faraday rotation, see Eq.~\eqref{eq:P_adiabat}.}

With decrease of the temperature the spectrum becomes asymmetric, which evidences the increasing role of non-commutativity of spin components. In the limit of \addDima{zero} temperature the spectrum contains the Stokes components only\addDima{. % in agreement with the discussion above.
%
% The dependence of the peak areas on their number allows one to extract the exchange interaction constant and effective spin temperature from the experimental spin noise and Raman spectra. To demonstrate this let us analyze the limiting cases for Eq.~\eqref{eq:An_pm}. For the strong exchange interaction, $\mu_\pm\gg 1$, we obtain
% \begin{equation}
%   \label{eq:An_strong}
%   A_n^\pm=\frac{\sqrt{\pi}(2n-1)!!}{\sqrt{\mu_++\mu_-}n!2^{n+1}}\left(\frac{\mu_\pm}{\mu_++\mu_-}\right)^n.
% \end{equation}
% For large $n$ this expression can be simplified as
% \begin{equation}
%   \label{eq:An_strong_large}
%   A_n^\pm=\frac{1}{2\sqrt{(\mu_++\mu_-)n}}\left(\frac{\mu_\pm}{\mu_++\mu_-}\right)^n.
% \end{equation}
% This expression shows, that the peaks' areas generally decay exponentially. However i
 In this limit $\mu_+=0$ and $A_n^-$ decay very slowly obeying the power law: %\commentDima{Is it the correct reference?}
  \begin{equation}
    \label{eq:An_sqrt}
    A_n^-\propto1/\sqrt{n}.
  \end{equation}
%The peaks at positive frequencies disappear completely, but at negative frequencies decay obeying the slow power law.
}

\subsection{Favorable conditions for the measurement of the high order correlators}
\label{sec:exp}

To successfully apply the resonance shift spin noise spectroscopy the exchange interaction should be quite strong. Indeed, in the opposite limit of weak \addDima{interaction, $\mu_\pm\ll1$, the} area of the $n$-th peak is given by Eq.~\eqref{eq:A_n_small} \addDima{and
% \begin{equation}
%   \label{eq:An0_gen}
%   A_n^{\pm}=\frac{(2n-1)!!}{2^{n+1}}\mu_\pm^n.
% \end{equation}
%One can see, that the areas decay exponentially and the area of the $n$th peak
is} proportional to $\omega_{ex}^{2n}$. %We give a more scrupulous analysis of this limit in Appendix~\ref{app:peaks}. 
Therefore, the exchange interaction should be strong, which is easily realized in semimagnetic semiconductors and many other systems, see Sec.~\ref{sec:conclusion}.

\begin{figure}
  \includegraphics[width=\linewidth]{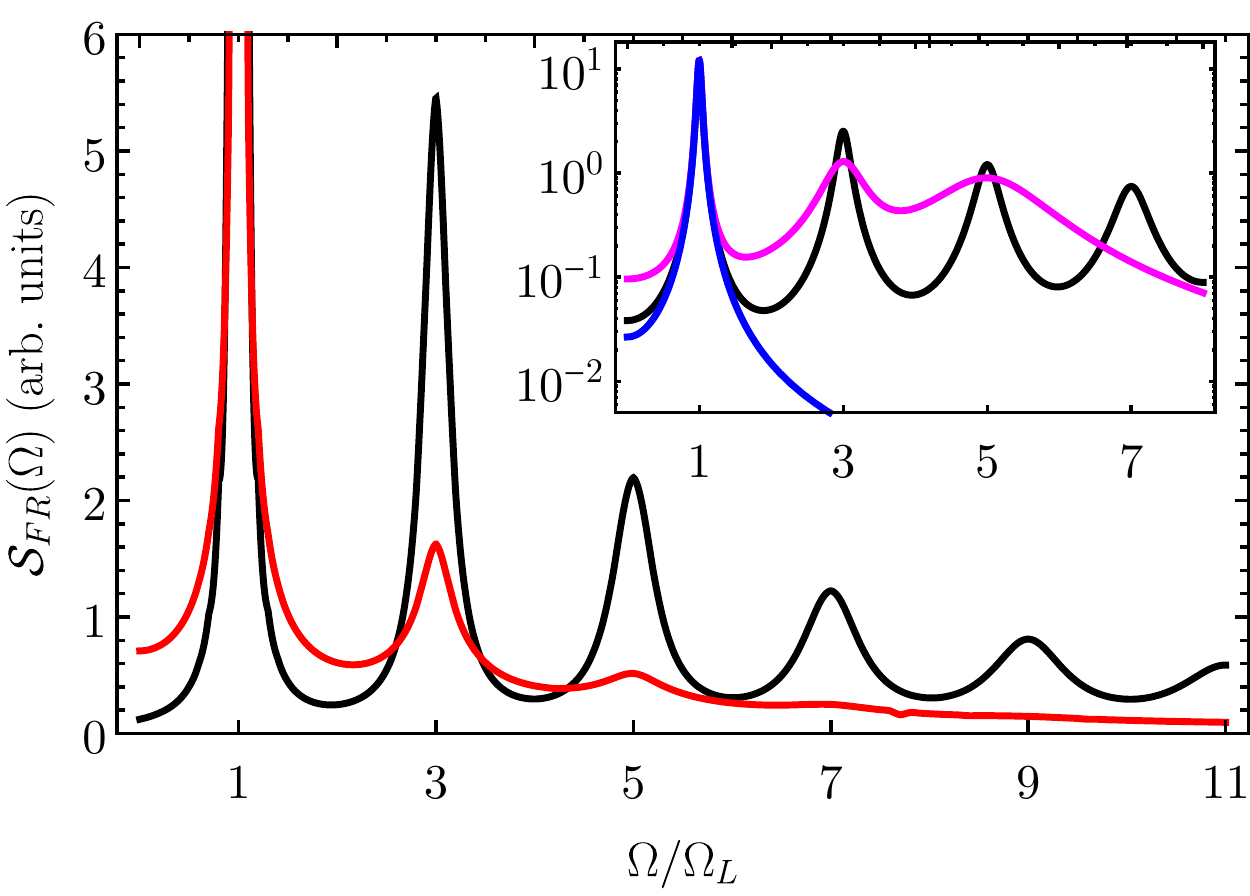}
  \caption{\addDima{Faraday} rotation noise spectra calculated after Eq.~\eqref{eq:spec_app} for the typical experimental parameters: $T=2$~K, $N=50$, $B_{exch}=\hbar\omega_{ex}/(\mu_B g)=1.5$~T with $g=2$, $\gamma=0.33$~meV, and $\Omega_L\tau_s=20$. The black and red curves corresponds to the strong  ($B=6$~T, $\Omega_L=168$~GHz) and moderate ($B=40$~mT, $\Omega_L=1.1$~GHz) external magnetic field, respectively. The inset compares Gaussian noise spectrum (black curve) \addDima{with} the Faraday rotation noise for a single spin $I=1/2$ (blue curve) \addDima{and} $I=5/2$ (magenta curve) in the limit $T\to 0$ and $\omega_{ex}\to\infty$ for $\Omega_L\tau_s=30$.
\commentDima{I made a new plot for $T=0$.}
}
  \label{fig:exp-nonnormal}
\end{figure}

Experimentally it is easier to measure the Faraday rotation noise spectra in the sub-GHz frequency range, which corresponds to the weak magnetic fields ${B\lesssim40}$~mT. In this case the average spin polarization is small, and we find
  \begin{equation}
    \braket{m(0)m(\tau)}=\frac{35N\omega_{ex}^2}{12\gamma^2}\cos(\Omega_L\tau)\e^{-|\tau|/\tau_s}.
  \end{equation}
The areas of the peaks decay quickly in this limit even despite the strong exchange \addDima{interaction.} The Faraday rotation noise spectrum for the typical experimental conditions for $B=40$~mT is shown in Fig~\ref{fig:exp-nonnormal} by red curve. One can see, that the peak at the frequency \addDima{$3\Omega_L$}
%, which describes the fourth order spin correlator,
 is much smaller, than the peak at $\Omega_L$, and the peak at $5\Omega_L$ is hardly visible in this limit.

More favorable conditions for the measurement of the high order spin correlations are realized in strong magnetic fields, when the spin polarization is large. The spin noise spectrum \addDima{for} $B=6$~T is shown by \addDima{the} blue curve in Fig.~\ref{fig:exp-nonnormal}, where one can distinctly see many peaks at the odd multiples of $\Omega_L$. In this case we obtain from Eq.~\eqref{eq:m2} the dimensionless spin correlation function
  \begin{equation}
    \braket{m(0)m(\tau)}=\frac{5N\omega_{ex}^2}{4\gamma^2}\e^{i\Omega_L\tau}\e^{-|\tau|/\tau_s}.
  \end{equation}
The areas of the peaks decay slowly in this case as \addDima{described} by Eq.~\eqref{eq:An_sqrt}. This limit corresponds to the dominance of the quantum spin fluctuations over the classical ones, and was not reached nor approached yet. To measure the Faraday rotation noise in the high frequency range one has to use special techniques, such as pulse trains~\cite{Berski-fast-SNS} or heterodyne detection~\cite{Heterodyne}. Nevertheless, we believe that this experimental challenge will be undertaken in the nearest future.

%\commentDima{With decrease of the temperature below $\hbar\Omega_L/k_B$, the quantum fluctuations begin to dominate over the thermal (or classical) ones. In the spin noise measurements performed to date the quantum limit of fluctuations was not reached nor approached yet. To observe zero-point {\Mn} spin fluctuations at temperatues $T\sim4$~K the applied magnetic field should be stronger, than $3$~T, which correspond to the Raman spin-flip measurements, rather than to the spin noise spectroscopy. However, detection of the quantum spin noise spectra can be done using a pulse trains~\cite{Berski-fast-SNS} or heterodyne detection~\cite{Heterodyne}.}

\section{Non-Gaussian spin noise}
\label{sec:abnormal}

In the previous sections we described the Faraday rotation noise spectra, and demonstrated, that they give access to the high order quantum spin correlation functions. However, under the assumption of many independent {\Mn} spins in the exciton localization volume, the high order spin correlation functions all can be reduced to the second order correlator. So it is interesting to go beyond this approximation and to study the non-Gaussian spin noise.

Generally, the noise spectrum is defined \addDima{by} Eq.~\eqref{eq:spin_spec}, where the Faraday rotation and the ellipticity angles are related to the exciton polarization by Eq.~\eqref{eq:thetas}. Using the averaged polarization correlation function~\eqref{eq:pys} we find the normalized Faraday rotation noise spectrum in the form
\begin{equation}
  \addDima{\mathcal S_{FR}(\Omega)}=\int\limits_{-\infty}^\infty\overline{\braket{p_y^*(0)p_y(\tau)}}_s\e^{\i\Omega\tau}\d\tau,
\end{equation}
where the dimensionless polarization is given by Eq.~\eqref{eq:Py_f_2}. In the adiabatic approximation, $\Omega_L\ll\gamma$, using the definition~\eqref{eq:J_def} we obtain
\begin{equation}
  p_y(t)=-\int\limits_{0}^\infty\e^{-\i\delta k-k}M(t)\d k\addDima{,}
\end{equation}
where we introduced the operator
  \begin{equation}
    M(t)=\sin[km(t)].
  \end{equation}
Then we perform averaging over the detuning, as defined in Eq.~\eqref{eq:pp}\addDima{,} and obtain the spectrum
\begin{equation}
  \label{eq:non-Gaussian_SN}
  \addDima{\mathcal S_{FR}(\Omega)}=\int\limits_{-\infty}^\infty\d\tau\e^{\i\Omega\tau}\int\limits_0^\infty\d k\e^{-2k}\braket{M(0)M(\tau)}_s.
\end{equation}
For non-Gaussian spin noise\addDima{,} the cumulants of $m(0)$ and $m(t)$ allow one to calculate this correlation function similarly to Sec.~\ref{sec:circ}.
%Using this expression the non-Gaussian spin noise spectrum can be calculated for the given spin $I$.

For example, in the presence of \addDima{the} resident charge carriers, {\Mn} spins are coupled with the carrier-mediated exchange \addDima{RKKY} interaction, which may eventually lead to \addDima{the} transition into the ferromagnetic phase~\cite{dietl:R3347}.
%For example, {\Mn} spins become strongly interacting and can not be considered as independent in 
For the {\Mn} concentration approaching the paramagnetic-ferromagnetic transition, their spins are no longer independent, and the spin noise in non-Gaussian. The spin fluctuations in this case can be described theoretically using the Landau theory~\cite{Stanley,Patashinskii}, effective polaron Hamiltonian~\cite{PhysRevLett.48.355,PhysRevLett.92.177203}, dynamical mean field theory~\cite{PhysRevB.73.075206,PhysRevB.99.045123}, or using more sophisticated approaches~\cite{PhysRevLett.89.277202,RevModPhys.90.025003}. In the vicinity of the phase transition the effective Larmor frequency decreases~\cite{soft_mode} and role of higher order cumulants increases~\cite{Sinitsyn-Correlators}.

However, in view of the application of the resonance shift spin noise spectroscopy to other systems we consider another situation. Namely, let us study the Faraday rotation noise induced by a single spin $\bm I$ ($N=1$) coupled to the optical resonance. In this case all the cumulants are equally important, see Eq.~\eqref{eq:cum_scaling}, so the spin noise is strongly non-Gaussian. This limit can be realized, \addDima{e.g.,} for deep impurities or atomic systems, see the next section.

For a single spin it is easier to calculate the Faraday rotation noise spectrum directly using the spin density matrix formalism, than using the cumulant expansion. We find the operator $M(t)$ from the equation of motion
\begin{equation}
  \label{eq:kinetic}
  \frac{\d M(\tau)}{\d\tau}=\frac{\i}{\hbar}\left[\mathcal H_0, M(\tau)\right]+\mathcal L\left\lbrace M(\tau)\right\rbrace,
\end{equation}
where $\mathcal H_0$ is defined in Eq.~\eqref{eq:H0} and $\mathcal L$ is the Lindblad operator, describing the spin relaxation. \addDima{Provided the transverse spin relaxation time $\tau_s$ is much shorter than the longitudinal one ($T_1$), we write the Lindblad operator in the form
\begin{equation}
  \mathcal L\left\lbrace M(\tau)\right\rbrace=\frac{1}{\tau_s}\left[2I_x M(\tau)I_x-I_x^2M(\tau)-M(\tau)I_x^2\right].
\end{equation}
The kinetic equation has a trivial initial condition $M(0)=\sin(k m)$, where $m$ is the Schrodinger operator defined in Eq.~\eqref{eq:m_def}. Finally, the correlation function in Eq.~\eqref{eq:non-Gaussian_SN} should be calculated using the steady state density matrix
\begin{equation}
  \rho=\left.\e^{-\mathcal{H}_0/(k_BT)}\middle/\Tr\left(\e^{-\mathcal{H}_0/(k_BT)}\right)\right..
  % \rho=\frac{\e^{-\mathcal{H}_0/T}}{\Tr\left(\e^{-\mathcal{H}_0/T}\right)}.
\end{equation}
}

As an example let us consider $I=1/2$. In this case
  \begin{equation}
    \sin(km)=2\sin\left(\frac{k\omega_{ex}}{2\gamma}\right)I_z.
  \end{equation}
Then the solution of Eq.~\eqref{eq:kinetic} simply reads
\begin{equation}
  M(\tau)=2\sin\left(\frac{k\omega_{ex}}{2\gamma}\right)\left[I_z\cos(\Omega_L\tau)+I_y\sin(\Omega_L\tau)\right]\e^{-\tau/\tau_s}.
\end{equation}
% \begin{multline}
%   \addDima{M(\tau)}=2\sin\left(\frac{k\omega_{ex}}{2\gamma}\right)\\\times\left[I_z\cos(\Omega_L\tau)+I_y\sin(\Omega_L\tau)\right]\e^{-\tau/\tau_s}.
% \end{multline}
\addDima{For any temperature the correlation function is the same:}
\begin{equation}
  \braket{M(0)M(\tau)}_s=\sin^2\left(\frac{k\omega_{ex}}{2\gamma}\right)\cos(\Omega_L\tau)\e^{-|\tau|/\tau_s}.
\end{equation}
Substituting this function in Eq.~\eqref{eq:non-Gaussian_SN} we find the non-Gaussian \addDima{Faraday rotation} noise spectrum for $I=1/2$:
\begin{equation}
  \label{eq:noise-12}
  \addDima{\mathcal S_{FR}(\Omega)}=\frac{1}{8}\frac{\omega_{ex}^2}{\omega_{ex}^2+4\gamma^2}\left[\mathcal P_1(\Omega)+\mathcal P_{1}(-\Omega)\right].
\end{equation}
This spectrum is shown by the blue curve in the Fig.~\ref{fig:exp-nonnormal}. Note, that its shape formally coincides with the spectrum of spin fluctuations, which is usually \addDima{measured} by the Pauli-blocking \addDima{spin} noise spectroscopy.

Generally, $\sin(km)$ can be presented as a linear combination of the operators $I_z^n$ with odd $n\le 2I$. As a result, the high order spin correlators can be reduced to a few lower orders, and the spectrum consists of a finite number of peaks. The maximum peaks' number is $n_{\rm{max}}=2[I-1/2]+1$, where square brackets denote the integer part. Similarly, in the multispin flip Raman spectra in circular polarization the maximum peaks' number is $2I$, which corresponds to the fact, that a single spin can not be flipped more than $2I$ times in one direction.

The Faraday rotation noise spectrum for \addDima{a single spin} $I=5/2$ \addDima{(in the limit $T=0$)} is shown in the inset in Fig~\ref{fig:exp-nonnormal} by a magenta curve. \addDima{One can see, that the maximum peaks' number is $5$, and the peaks are much broader, than for the Gaussian spin noise. This indicates, that the higher order spin correlators generally contain more information than the common second order one.}

%====================================================
\section{Discussion and conclusion}
\label{sec:conclusion}

To detect the higher order spin correlators\addDima{,} the ratio between the exchange broadening of the exciton resonance $\sqrt{N}\omega_{ex}$ should be comparable to or larger than the homogeneous linewidth $\gamma$, as shown in Sec.~\ref{sec:exp}. This condition is easily satisfied in Mn-doped QWs~\cite{Stuhler1995} and QDs~\cite{Kozyrev2019}, where up to $15$ peaks in the Raman \addDima{spin flip} spectra are visible. For these structures $\sqrt{N}\omega_{ex}\sim 2$~meV and $\gamma\sim 1$~meV.

For Mn-doped nanosystems, the number of the probed spins is typically very large $N\gtrsim 100$, and the total spin noise is almost Gaussian. However with increase of the {\Mn} concentration a limited number of closely located pairs of magnetic atoms appears. The strength of the exchange interaction in a pair can be of the order of $0.5$~meV~\cite{PhysRevB.33.1789,PhysRevB.96.241303,Cherbunin-pairs}, and the ground state of the pair is the singlet spin state. The difference between the interaction constants with the heavy hole in the localized exciton for the two spins in a pair leads to the mixing between singlet and triplet states. We expect, that it will manifest itself as another comb of peaks with the frequencies a bit larger than $n\Omega_L$ in Faraday rotation noise spectrum. Due to the small number of pairs their contribution is non-Gaussian, and therefore contains detailed information about the spin dynamics of the pair of strongly coupled spins.

\addDima{Importantly, the} resonance shift quantum spin noise spectroscopy can be applied to a very broad class of spin systems. For example, it can be applied to measure the nuclear spin fluctuations in QDs. In this case the main requirement for the detection of high order spin \addDima{correlators} is the sizable hyperfine interaction strength as compared \addDima{with} the inverse lifetime of the excited state. For example, in GaAs-based QDs the hyperfine interaction constant for electrons is $A\approx 100~\mu$eV~\cite{Chekhovich_constants,book_Glazov}, so for small QDs with $N\sim 10^4$ one has $\sqrt{N}\omega_{ex}\sim A/\sqrt{N}\sim1~\mu$eV, which is the typical exciton homogeneous linewidth~\cite{warburton2013}. Thus we expect, that the higher order nuclear spin correlators can be measured for small QDs as well as for the small colloidal nanocrystals.%~\cite{Beaulac973}.

Resonance shift spin noise spectroscopy is particularly useful, when the number of probed spins is small, because in this case the high order spin correlators can not be reduced to the lower orders. To measure these correlation functions, the spin-related (e.g. hyperfine) structure of individual optical transitions should be visible. There are many examples of such systems: For NV$^-$ centers in diamond the hyperfine interaction constant with the nearest $C^{13}$ atom \addDima{can reach} $0.1~\mu$eV~\cite{Smeltzer_2011}, which is approximately two times larger, than the homogeneous linewidth at liquid helium temperatures~\cite{PhysRevLett.97.083002}. For rare earth ions, $A$ can reach $10~\mu$eV, while the homogeneous linewidth is a few times smaller~\cite{Macfarlane1987}. In the past few years the van der Waals heterostructures are under intense investigation. For localized spatially indirect excitons, the hyperfine \addDima{interaction} induced spin relaxation time is predicted to be $T_2^*\sim\hbar/(\sqrt{N}\omega_{ex})\sim1$~ns~\cite{MX2_hyperfine, MX2_Avdeev}, which is an order of magnitude shorter, than the exciton lifetime $1/\gamma=10$~ns~\cite{IX_MX2}, so \addDima{the} nuclear related broadening of the optical transition exceeds its linewidth by an order of magnitude thanks to the small exciton Bohr radius. The similar situation is also realized for the lead halide perovskites where $T_2^*$ is comparable with $1/\gamma$~\cite{belykh2019coherent}. So these systems are prominent for the resonance shift nuclear spin noise spectroscopy. Apart from the solid state physics, the hyperfine structure of optical transitions is quite routinely observed for atoms, such as K, Na, Rb, Cs; and for simple molecules, such as I$_2$~\cite{Xu2000}.%,Dube:04},
Therefore these systems are also promising for the resonance shift spin noise spectroscopy.

In conclusion, we have developed a theory of a class of optical phenomena that occur in optically transparent solids with localized spins (e.g. {\Mn} spins in diluted magnetic semiconductors), forming a basis for a set of experimental methods, which can be generically called resonance shift spin noise spectroscopy. The distinctive feature of these phenomena is that the spins do not directly participate in the probed optical transitions (e.g. excitonic ones), but they shift such transitions via the spin-spin interactions. To demonstrate the universality and power of this approach, we obtained the expressions for multispin flip Raman spectra in diluted-magnetic quantum wells and calculated the Faraday rotation noise spectra. We predict multiple overtones of the Larmor frequency in the spectra, which reflect the contributions of the \addDima{high order} correlation functions of the spin fluctuations. Our predictions open a way for the experimental investigation of \addDima{high order} spin noise, including quantum noise. Our approach is directly extendable to a wide range of solid-state and atomic systems.

\begin{acknowledgments}
We gratefully acknowledge the fruitful discussions with M. M. Glazov and D. Scalbert.
\addDima{D.S.S. was supported by the Russian Science Foundation Grant No. 19-72-00081 and the Basis Foundation.}
% D.S.S. was partially supported by the Russian Science Foundation Grant No. 19-72-00081, RF President Grant No. MK-1576.2019.2, Russian Foundation for Basic Research (Grant No. 17-02-0383) and the Basis Foundation.
K.V.K. was supported by a grant from Saint-Petersburg State University and DFG ID 39411635 (Project No.40.65.62.2017).
\end{acknowledgments}

% \begin{figure}
%   \includegraphics[width=\linewidth]{Exp_spec}
%   \caption{The Faraday rotation noise spectra calculated after Eq.~\eqref{eq:spec_app} for the typical experimental conditions: $\gamma=0.33$~meV, $T=2$~K, $N=50$, $B_{exch}=\hbar\omega_{ex}/(\mu_B g)=1.5$~T with $g=2$. The blue curve corresponds to strong external magnetic field $B=6$~T ($\Omega_L=168$~GHz) with $\Omega_L\tau_s=40$ and the red curve --- to weak magnetic field $B=40$~mT ($\Omega_L=1.1$~GHz) with $\Omega_L\tau_s=4$.}
%   \label{fig:exp-supp}
% \end{figure}

% \begin{figure}
%   \includegraphics[width=\linewidth]{spec52-5.pdf}
%   \caption{Faraday rotation noise spectra for $T=\infty$ and $m=\infty$ calculated for a single spin $I=1/2$ (black curve), $I=5/2$ (red curve) and for Gaussian spin noise (blue curve).}
%   \label{fig:non-normal-supp}
% \end{figure}

\appendix

\section{Calculation of the polarization}
\label{app:P}

We start from the general form of the Hamiltonian~\eqref{eq:H}. Its Hilbert space is a direct product of the states of the spin system and the excitonic states including exciton vacuum state. For heavy hole excitons there are four states, which can be labeled by the electron spin projection $S_z^e=\pm1/2$ and the hole spin $S_z^h=\pm 3/2$. The excitonic states can be denoted as $\ket{k}$, where $k=1,2,\ldots$. In the first order in the incident field amplitude, one can consider a single exciton states only, so the exciton Hamiltonian has the form
\begin{equation}
  \mathcal H_{exc}=\sum_k \mathcal H_{exc}^{k'k} c_{k'}^\dag c_{k},
\end{equation}
where $c_k$ ($c_k^\dag$) are the annihilation (creation) operators for the states $\ket{k}$. This Hamiltonian describes the fine structure of the excitonic levels and exciton interaction with the external magnetic field.

The coherent exciton generation is described by Eq.~\eqref{eq:V}, where
\begin{equation}
  \bm P=\sum_k \bm d_k c_k
\end{equation}
with $\bm d_k$ being the dipole moments of the excitonic states. In the particular model, which is used in the derivation of Eq.~\eqref{eq:Pt}, there are two excitonic states with the dipole moments $\bm d_{\pm}=d(-\bm e_x\mp\i\bm e_y)/\sqrt{2}$, where $\bm e_\alpha$ are the unit vectors along the corresponding axes.

The Hamiltonian $\mathcal H_0$ describes the magnetic spin system only and does not contain operators $c_k$ and $c_k^\dag$. Moreover, the Hamiltonian of the spin-exciton exchange interaction has the form
\begin{equation}
  \mathcal H_{int}=\sum_{k,k',i}\bm I_{i}\mathcal H_{int}^{kk'}c_k^\dag c_{k'}.
\end{equation}
Note, that this Hamiltonian contains \addDima{off diagonal} terms (with $k\neq k'$) and coincides with Eq.~\eqref{eq:H_int}.

The \addDima{operator of the system evolution} is
\begin{equation}
  U=\mathcal T\exp\left[-\frac{\i}{\hbar}\int_0^t\mathcal H(t')\d t'\right],
\end{equation}
and the Heisenberg polarization operator is
\begin{equation}
  \bm P^{(0)}(t)=U^\dag\bm P U.
\end{equation}
We are interested in the contribution $\bm P(t)$ to $P_\alpha^{(0)}(t)$ \addDima{only}, which is linear in the amplitude of the probe light $\bm E$. This operator acts only in the Hilbert space of the exciton vacuum state, so it is given by
\begin{multline}
  \bm P(t)=\e^{\frac{\i}{\hbar}\mathcal H_0 t}\bm P\int\limits_0^t\e^{-\frac{\i}{\hbar}(\mathcal H_0+\mathcal H_{exc}+\mathcal H_{int})\tau}\\
  \times\frac{\i}{\hbar}\left(\bm P^\dag\bm E\right)\e^{-\frac{\i}{\hbar}\mathcal H_0(t-\tau)}\d\tau.
\end{multline}
Since the spin Hamiltonian $\mathcal H_0$ commutes with $\bm P$, this expression can be written as
\begin{equation}
  \label{eq:P_Phi}
  \bm P(t)=\frac{\i}{\hbar}\bm P\int\limits_0^t\Phi(t,\tau)\d\tau\left(\bm P^\dag\bm E\right),
\end{equation}
where
\begin{equation}
  \Phi(t,\tau)=\e^{\frac{\i}{\hbar}\mathcal H_0 t}\e^{-\frac{\i}{\hbar}(\mathcal H_0+\mathcal H_{exc}+\mathcal H_{int})\tau}\e^{-\frac{\i}{\hbar}\mathcal H_0(t-\tau)}.
\end{equation}
To simplify this expression we note that
\begin{equation}
  \frac{\partial\Phi(t,\tau)}{\partial\tau}=-\frac{\i}{\hbar}\Phi(t,\tau)\left[\mathcal H_{exc}+\tilde{\mathcal H}_{int}(t-\tau)\right]
\end{equation}
with $\tilde{\mathcal H}_{int}(t)$ given by Eq.~\eqref{eq:Hint_int}. One can readily see, that $\Phi(t,0)=1$, so the solution of this equation is
\begin{equation}
  \Phi(t,\tau)=\mathcal T\exp\left\{-\frac{\i}{\hbar}\int\limits_0^\tau\left[\mathcal H_{exc}+\tilde{\mathcal H}_{int}(t-\tau')\right]\d\tau'\right\}.
\end{equation}
Substituting this expression in Eq.~\eqref{eq:P_Phi} we see, that $\bm P(t)=\bm P\psi$, where
\begin{multline}
  \psi=\frac{\i}{\hbar}\int\limits_0^t
  \mathcal T\exp\left\{-\frac{\i}{\hbar}\int\limits_0^\tau\left[\mathcal H_{exc}+\tilde{\mathcal H}_{int}(t-\tau')\right]\d\tau'\right\}\d\tau\\
  \times\left(\bm P^\dag\bm E\right)
\end{multline}
in agreement with Eq.~\eqref{eq:P_gen}. One can readily check, that it satisfies Eq.~\eqref{eq:psi} indeed.

\section{Cumulant expansion}
\label{app:cumulants}

The generating function for the cumulants of $\mathcal J(0)$ and $\mathcal J(\tau)$ can be \addDima{taken in the following form:}
  \begin{equation}
    \mathcal K=\ln\braket{\e^{\i\alpha\mathcal J(0)-\i\beta\mathcal J(\tau)}}.
  \end{equation}
Its Tailor series defines the cumulants $\kappa\left(\mathcal J^l(0),\mathcal J^{(n-l)}(\tau)\right)$ as
\begin{equation}
  \mathcal K =\sum_{n=1}^\infty\sum_{l=0}^n
%C_n^l
{{n}\choose{l}}
%\left( \right)
\alpha^l(-\beta)^{n-l}\kappa\left(\mathcal J^l(0),\mathcal J^{(n-l)}(\tau)\right)\addDima{.}
\end{equation}
%where ${{n}\choose{l}}$ is the binomial coefficient.
Comparing this expression with the correlator in Eq.~\eqref{eq:pp} we find
\begin{multline}
  \label{eq:JJ_gen}
  \braket{\left[\overline{\mathcal T}\e^{\i\mathcal J(0)}\right]\left[\mathcal T\e^{-\i\mathcal J(\tau)}\right]}
%\\
=\exp\left\{\sum_{n=1}^\infty\sum_{l=0}^{2n}\frac{(-1)^{n+l}}{(2n-l)!l!}
    \right.\\\left.\times
\kappa\left(\mathcal J^l(0),\mathcal J^{(2n-l)}(\tau)\right)\right\},
\end{multline}
where we took into account that the cumulants of the odd orders vanish in the absence of \addDima{the} spin polarization along the $z$ axis ($\braket{m(t)}=0$). The cumulants of the operators should be calculated using the normal time ordering for the powers of $\mathcal J(\tau)$, reverse time ordering for $\mathcal J(0)$ and putting $\mathcal J(0)$ always to the left of $\mathcal J(\tau)$.

To simplify the following, we assume, that the {\Mn} spins, $\bm I_i$ in Eq.~\eqref{eq:I_tot}, are independent. In this case $\mathcal J(t)$, as defined in Eq.~\eqref{eq:J_def} also consists of $N$ independent contributions $\mathcal J_i(t)\sim1/N$. Then a cumulant of the sum of independent variables takes the form~\cite{Gardiner}
  \begin{equation}
    \kappa\left(\mathcal J^l(0),\mathcal J^{(2n-l)}(\tau)\right)=\sum_{i=1}^N\kappa\left(\mathcal J_i^l(0),\mathcal J_i^{(2n-l)}(\tau)\right).
  \end{equation}
  From this relation one can see the scaling law for the cumulants
  \begin{equation}
    \label{eq:cum_scaling}
    \kappa\left(\mathcal J^l(0),\mathcal J^{(2n-l)}(\tau)\right)\propto1/N^{2n-1}.
  \end{equation}
  The larger is $N$ the less important are the cumulants of the high orders.

In the limit of many independent {\Mn} spins, $N\gg1$, one can neglect all the cumulants except for $n=1$ (the second order one). This corresponds to the normal or Gaussian spin noise. In this case Eq.~\eqref{eq:JJ_gen} reduces to
\begin{equation}
  \label{eq:Gaussian}
  \braket{\overline{\mathcal T}\e^{\i\mathcal J(0)}\mathcal T\e^{-\i\mathcal J(\tau)}}
  =\exp\left(\braket{\mathcal J(0)\mathcal J(\tau)}-\braket{\mathcal J^2(0)}_s\right),
\end{equation}
and from Eq.~\eqref{eq:pp} we obtain
\begin{equation}
  \label{eq:pp_int_J}
  \overline{\braket{p_+^*(0)p_+(\tau)}}=
  \int\limits_0^\infty\e^{-2k+{\braket{\mathcal J(0)\mathcal J(\tau)}-\braket{\mathcal J^2(0)}_s}}
  \d k.
\end{equation}

The second order correlator of $\mathcal J(t)$ can be presented as a double integral using its definition~\eqref{eq:J_def}. The correlation function $\braket{m(t_1)m(t_2)}$ depends on $t_1-t_2$ \addDima{only}, so the double integral can be reduced to a single integral as follows:
  \begin{subequations}
    \label{eq:J2}
    \begin{equation}
      \label{eq:J2tau}
      \braket{\mathcal J(0)\mathcal J(\tau)}=\int\limits_{-k}^k(k-|k'|)\braket{m(0)m(\tau+k'/\gamma)} \d k',
    \end{equation}
    \begin{equation}
      \label{eq:J20}
      \braket{\mathcal J^2(0)}_s=2\int\limits_0^k(k-k')\braket{m(0)m(k'/\gamma)}_s \d k'.
    \end{equation}
  \end{subequations}
Substitution of these expressions in Eq.~\eqref{eq:pp_int_J} yields the polarization correlation function, which defines the Raman spin flip spectrum, see \addDima{Eq.~\eqref{eq:S_tilde}}.

\section{Areas of the peaks}
\label{app:peaks}

In the realistic limit $\tau_s\gamma\gg 1$ using Eq.~\eqref{eq:Iz2} we obtain from Eq.~\eqref{eq:J2}
  \begin{equation}
    \braket{\mathcal J(0)\mathcal J(\tau)}=\left(\mathcal J_+\e^{-\i\Omega_L\tau}+\mathcal J_+\e^{\i\Omega_L\tau}\right)\e^{-|\tau|/\tau_s}
  \end{equation}
and $\braket{\mathcal J^2(0)}_s=\braket{\mathcal J^2(0)}$ with
\begin{equation}
    \label{eq:Jpm}
  \mathcal J_\pm=\frac{\omega_{ex}^2}{\Omega_L^2}\left(\braket{I_{z}^2}\mp\frac{\braket{I_{x}}}{2}\right)\left[1-\cos\left(\frac{\Omega_L}{\gamma}k\right)\right].
\end{equation}
Substituting these expressions in Eq.~\eqref{eq:pp_int_J} and decomposing the exponent into series like in Eq.~\eqref{eq:corr_power} we find
\begin{widetext}
\begin{equation}
  \label{eq:spec_app}
  \overline{\braket{p_+^*(0)p_+(\tau)}}=\sum_{n=0}^{\infty}\sum_{l=0}^n\addDima{\frac{\e^{\i(2l-n)\Omega_L\tau-n|\tau|/\tau_s}}{l!(n+l)!}}\int\limits_0^\infty\mathcal J_+^l\mathcal J_-^{n-l}\e^{-2k-\mathcal J_+-\mathcal J_-}\d k.
\end{equation}
%The term with $n=0$ and $l=0$ here should be dropped

In the adiabatic approximation ($\Omega_L\ll\gamma$) Eq.~\eqref{eq:Jpm} reduces to $\mathcal J_\pm=k^2\mu_\pm$, so from Eq.~\eqref{eq:spec_app} we obtain
\begin{equation}
  \overline{\braket{p_+^*(0)p_+(\tau)}}=\sum_{n=0}^{\infty}\sum_{l=0}^n\addDima{\frac{\mu_+^l\mu_-^{n-l}}{l!(n+l)!}}\e^{\i(2l-n)\Omega_L\tau-n|\tau|/\tau_s}\int\limits_0^\infty\e^{-2k-k^2(\mu_++\mu_-)}k^{2n}\d k.
\end{equation}
% \begin{multline}
%   \overline{\braket{p_+^*(0)p_+(\tau)}}=\int\limits_0^\infty\e^{-2k}\e^{-k^2(\mu_++\mu_-)}\sum_{n=0}^{\infty}k^{2n}\\
%   \times\sum_{l=0}^n\frac{l!(n+l)!}{(n!)^2}\mu_+^l\mu_-^{n-l}\e^{\i(2l-n)\Omega_L\tau}\e^{-n|\tau|/\tau_s}\addDima{\d k}.
% \end{multline}
In the next step we introduce $n'=n-2l$ and $l'=l$ for $n'\ge 0$ and $l'=n-l$ otherwise. Then we rewrite this expression as
\begin{equation}
  \overline{\braket{p_+^*(0)p_+(\tau)}}=\sum_{n'=-\infty}^{\infty}
  \sum_{l'=0}^\infty\addDima{\frac{\mu_{\sign(n')}^{l'}\mu_{-\sign(n')}^{|n'|+l'}}{l'!(|n'|+l')!}}
  \e^{-\i n'\Omega_L\tau-(|n'|+2l')|\tau|/\tau_s}
\int\limits_0^\infty\e^{-2k-k^2(\mu_++\mu_-)}k^{2(|n'|+2l')}
  \d k,
\end{equation}
% \begin{multline}
%   \overline{\braket{p_+^*(0)p_+(\tau)}}=\sum_{n'=-\infty}^{\infty}\e^{-\i n'\Omega_L\tau}\int\limits_0^\infty\e^{-2k}\e^{-k^2(\mu_++\mu_-)}\\
%   \times\sum_{l'=0}^\infty k^{2(|n'|+2l')}\frac{l'!(|n'|+l')!}{[(|n'|+2l')!]^2}\mu_{\sign(n')}^{l'}\mu_{-\sign(n')}^{|n'|+l'}\\
%   \times\e^{-(|n'|+2l')|\tau|/\tau_s}\addDima{\d k},
% \end{multline}
where we assume $\sign(0)\equiv1$ to be specific. Finally, the Fourier transform of this expression [see Eq.~\eqref{eq:S_tilde}] \addDima{yields}
%\begin{widetext}
\begin{equation}
  \mathcal S_{++}(\Omega)=\sum_{n\neq 0}\sum_{l=0}^\infty\int\limits_0^\infty\e^{-2k}\e^{-k^2(\mu_++\mu_-)}k^{2(|n|+2l)}%\\\times
  \addDima{\frac{\mu_{\sign(n)}^{l}\mu_{-\sign(n)}^{|n|+l}}{l!(|n|+l)!}}
 %\frac{2\tau_s/(|n|+2l)}{1+\left[(\Omega-n\Omega_L)\tau_s/(|n|+2l)\right]^2}\d k,
 \mathcal P_{n,|n|+2l}(\Omega)\d k,
\end{equation}
\end{widetext}
where we omitted the primes for the brevity, neglected the terms with $n=0$, and introduced the Lorentzian functions
\begin{equation}
  \mathcal P_{n,n'}(\Omega)=\frac{2\tau_s/n'}{1+\left[(\Omega-n\Omega_L)\tau_s/n'\right]^2},
\end{equation}
which describe the peaks at \addDima{the} frequencies $\pm n\Omega_L$ with the widths $n'/\tau_s$.

\bibliography{Multispin}
%\bibliography{all-1}
\end{document}